\documentclass[12pt,draftclsnofoot,onecolumn]{IEEEtran}
\usepackage[colorlinks]{hyperref}
\usepackage{amssymb}
\usepackage{algorithm}
\usepackage{algorithmic}
\usepackage{xcolor}
\usepackage{mathrsfs}
\usepackage{blkarray}
\usepackage{amsthm}
\usepackage{amsmath}
\usepackage{hyperref}

\ifCLASSINFOpdf
  \usepackage[pdftex]{graphicx}
  \graphicspath{{./figures/}}
  \DeclareGraphicsExtensions{.pdf,.jpeg,.png}
\else
\fi

\usepackage{amsfonts}
\usepackage{amsmath}
\usepackage{caption}
\usepackage{subcaption}
\usepackage[inline]{enumitem}

\usepackage{algorithm}
\usepackage{algorithmic}
\usepackage{xcolor}
\usepackage{mathrsfs}
\usepackage{blkarray}
\usepackage{amsthm}

\newtheorem{theorem}{Theorem}
\newtheorem{definition}{Definition}

\newtheorem{lemma}{Lemma}

\usepackage{circledsteps}

\ifCLASSINFOpdf
\else
\fi
\usepackage{amsmath}
\usepackage{amsfonts}

\begin{document}
\title{Reflection Map Construction: Enhancing and Speeding Up Indoor Localization}
\author{Milad~Johnny,~\IEEEmembership{Member,~IEEE,}
        Shahrokh Valaee,~\IEEEmembership{Fellow,~IEEE,}
\thanks{The authors are with the Department of Electrical and Computer Engineering,
University of Toronto, Toronto, ON M5S, Canada (e-mail: milad.johnny@utoronto.ca; valaee@ece.utoronto.ca).}}% <-this % stops a space
%\thanks{J. Doe and J. Doe are with Anonymous University.}
\maketitle
\begin{abstract}
This paper introduces an indoor localization method using fixed reflector objects within the environment, leveraging a base station (BS) equipped with Angle of Arrival (AoA) and Time of Arrival (ToA) measurement capabilities. The localization process includes two phases. In the offline phase, we identify effective reflector points within a specific region using significantly fewer test points than typical methods. In the online phase, we solve a maximization problem to locate users based on BS measurements and offline phase information. We introduce the reflectivity parameter (\(n_r\)), which quantifies the typical number of first-order reflection paths from the transmitter to the receiver, demonstrating its impact on localization accuracy. The log-scale accuracy ratio (\(R_a\)) is defined as the logarithmic function of the localization area divided by the localization ambiguity area, serving as an accuracy indicator. We show that in scenarios where the Signal-to-Noise Ratio (SNR) approaches infinity, without a line of sight (LoS) link, \(R_a\) is upper-bounded by \(n_r \log_{2}\left(1 + \frac{\mathrm{Vol}(\mathcal{S}_A)}{\mathrm{Vol}(\mathcal{S}_{\epsilon}(\mathcal{M}_s))}\right)\). Here, \(\mathrm{Vol}(\mathcal{S}_A)\) and \(\mathrm{Vol}(\mathcal{S}_{\epsilon}(\mathcal{M}_s))\) represent the areas of the localization region and the area containing all reflector points with a probability of at least \(1 - \epsilon\), respectively.
\end{abstract}


\IEEEpeerreviewmaketitle
\section{Introduction}
%\linespread{1.5}
In recent years, indoor positioning systems have garnered significant attention due to their diverse range of applications, including asset tracking, location-based services, and navigation assistance in complex indoor environments, such as shopping malls, airports, and industrial facilities \cite{ZhuXiaoqiang}, \cite{JangBeakcheol}. There exist many different approaches for indoor localization. Time of Arrival (ToA) based positioning relies on measuring the time taken for a wireless signal to travel from a transmitter to a receiver. This method typically requires knowledge of the locations of at least three base stations during the positioning process. However, it imposes requirements for time synchronization and the existence of a line of sight (LoS) between the transmitter and receiver, which can be challenging to achieve in practice. Another group of approaches use Angle of Arrival (AoA) which involves measuring the incident angle of a \color{black}radio \color{black} signal. For accurate results, the receiver typically needs an antenna array. However, for complex indoor environments without LoS, the localization strategy is exceptionally challenging. Another technique for indoor localization is based on the propagation loss model, this approach attempts to establish a mathematical model that relates signal strength to the distance between the transmitter and receiver. However, the presence of multipath effects in indoor environments can make it difficult to create an accurate propagation loss model. The \color{black}positioning \color{black} fingerprint method is gaining popularity in indoor positioning technology. This method relies on creating a large database of \color{black} received signal-strength (RSS) \color{black} fingerprints at various known locations within the indoor environment. When a user's device measures signal strengths, it can compare these measurements to the fingerprints in the database to estimate its location.

In summary, while various indoor localization methods exist, the position fingerprint method is gaining prominence due to its practical advantages, making it the preferred choice for many indoor positioning applications. However, achieving accurate and reliable indoor localization in these settings remains a formidable challenge due to the huge number of measurements, and complex searching algorithm through the saved data set. \color{black} Additionally, even minor changes in the carrier frequency within the data communication link can cause fluctuations in the RSS parameter from the saved dataset, making implementation difficult. \color{black} Similarly, machine learning techniques have demonstrated remarkable performance in indoor localization, particularly when trained on extensive datasets. However, they often face practical challenges when transitioning from controlled, training-oriented environments to real-world deployment scenarios. One of the primary hurdles is the resource-intensive process of assembling and maintaining large and diverse fingerprinting datasets \cite{WangBang, ZhongzeZhang, LiQiao2, ZhaoYunming, LiQiao}. This process involves meticulously collecting and labelling data for various user locations in diverse indoor settings, which can be both time-consuming and costly \cite{ZhaoPing, ChenPan, HeZheLi, SongQianwen, Berruet, Montella0, HanandHuang}. In \cite{drvalaee4}, the authors demonstrate that the Compressed Sensing (CS) scheme can be employed in the offline phase to reconstruct the radio map based on a small number of \color{black}RSS \color{black} measurements. This technique relies on the assumption that RSS readings exhibit smooth variations across the area and that its Fourier coefficients possess a sparse nature. However, this assumption may not hold in a complex environment with many reflectors. In addition, the complexity of the measurement process remains high. In \cite{Tahsin}, the authors address a solution for the sparse inverse problem using a binary programming approach for localizing a transmitter in an indoor multipath environment but the approach is still time-consuming and needs LoS between transmitters and receiver. To address these multifaceted challenges, this paper introduces a localization strategy that leverages fixed reflection \color{black}objects \color{black} within the indoor environment. The goal is to provide precise user positioning and enhance the efficiency of indoor positioning systems. 
\begin{figure}
  \centering
  \includegraphics[width=0.5\textwidth]{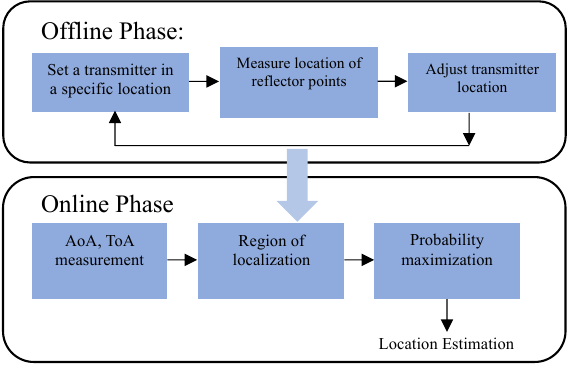}
  \caption{Block diagram of proposed indoor localization strategy}
  \label{Blockdia}
\end{figure}
Referring to Fig. \ref{Blockdia}, our approach consists of \color{black} two phases\color{black}:\\
{\it i) Topology Mapping and Data Collection (Offline Phase):} The first \color{black} phase \color{black} involves identifying the locations of effective fixed reflector points. What sets our approach apart is its efficiency in data collection; we demonstrate how these critical pieces of information can be obtained with significantly fewer test points compared to fingerprint-based strategies reliant on radio map reconstruction. This stage can be accomplished by deploying a transmitter with a known location, capable of moving in the environment while measuring signal parameters such as ToA and AoA at the BS. With ToA and AoA measurements accessible from the transmitter's various locations, the BS can ascertain the positions of a subset of activated reflector points. These limited data points can then be used to construct a more comprehensive map representing the effective fixed reflector layout.

{\it ii) User Localization (Online Phase):} The second step involves estimating the user's location by maximizing a function that incorporates information obtained in the previous stage. At the BS, when evaluating a candidate position for the user's location, we can \color{black} map a \color{black} set of 2-tuple measurements (AoA, ToA) \color{black}to the \color{black} potential reflection positions. By comparing the positions of these points with our fixed reflector map, we can establish an evaluation criterion. To pinpoint the optimal user location, we formulate a maximization problem that accounts for the likelihood of each set of fixed reflectors. This holistic approach ensures that every effective reflector in the environment is appropriately considered in the localization process.

{\it Organization:} In Section II, we present the system model and problem formulation. Section III delves into our strategy for identifying and reconstructing the reflection map, accompanied by relevant terms and definitions. In Section IV, we present our reflection points construction from a limited number of test points. In Section V, we elaborate on our localization algorithm, introducing the Log-scale accuracy ratio and calculating its upper bound as the signal-to-noise ratio ($\mathrm{SNR}$) tends to infinity. This measure, distinct from common metrics such as the Cramer-Rao lower bound (CRLB) on the Mean Squared Error (MSE), as well as global Minimum Mean Squared Error (MMSE) bounds like the Ziv-Zakai bound (ZZB) \cite{Bell, Bilik} and the Weiss-Weinstein bound (WWB) \cite{Weinstein}, is essentially influenced by environmental conditions for Non-LoS connections. Besides, we introduce a rapid pre-localization algorithm followed by a more precise localization technique. Section VI provides an overview of our simulation results, and finally, we conclude the paper in Section VII. \\     
\textbf{Notation:} In this paper, random variables are shown in capital letters and their realizations are shown in lowercase letters. Sets are shown in a calligraphic font and we use $|\mathcal{X}|$ to show the cardinally of the set $\mathcal{X}$. The probability of event $A$ is denoted by $\mathbb{P}\left({A}\right)$. The $\lvert{{\bf v}}\rvert$ represents the magnitude of the vector $\bf{v}$. The $\|{.}\|_2$ represents the Euclidean norm and $\mathbb{E}\{.\}$ indicates expected value.

%%%%%%%%%%%%%%%%%%%%%%%%%%%%%%%%%%%%%%%%%%%%%%%%%%%%%%%%%%%%%%%%%%%%%%%%%%%%%%%%%%
%%%%%%%%%%%%%%%%%%%%%%%%%%%%%%%%%%%%%%%%%%%%%%%%%%%%%%%%%%%%%%%%%%%%%%%%%%%%%%%%%%%%%% 
\section{System Model}
\label{sec:system_model}
In this paper, we focus on mm-waves, which reflect on most of the surfaces in the environment and behave quasi-optically. Many measurement results confirm that for high-frequency applications the strongest part of the multipath components can be determined based on ray tracing techniques \cite{Aryanfar, Dagefu, Lecci, ChoiHyuckjin, YangJie}. For the sake of simplicity, we are considering a two-dimensional environment in which we focus on the localization of a single mobile user. The generalization to a three-dimensional environment is straightforward. We assume the presence of a BS with a known location, acting as a reference point. 
Without loss of generality, we assume the presence of a single user, extension to multiple users is straightforward. Throughout all sections of this paper, we adopt the following notation to represent various locations of significant objects within our environment. The two tuples ${\bf p}_{b}=(x_b, y_b)$ and ${\bf p}_{u}=(x_{u}, y_{u})$ represent the BS and user locations, respectively. We denote ${\bf p}_{s_i}=(x_{s_i}, y_{s_i}) \in \mathcal{M}_{s}$, \color{black} $0 \leq \lvert{\mathcal{M}_{s}}\rvert< \infty$ \color{black}as the location of the $i$-th fixed reflector point. These points may include scatterers or specific points on surfaces like walls, collectively constituting a model for environmental reflectors. Given the real-world context, the set $\mathcal{M}_{s}$ is continuous and encompasses an infinite number of elements. 

We assume that the BS has an antenna with many elements and it can measure the set $\mathcal{M}_e$ defined as follows:
\begin{align}
\mathcal{M}_e&=\{{M_i}\}, 1\leq i \leq \vert{\mathcal{M}_e}\vert \nonumber\\
&=\{ \left({\theta_1,\tau_1}\right),\dots,\left({\theta_{\vert{\mathcal{M}_e}\vert},\tau_{\vert{\mathcal{M}_e}\vert}}\right)\},
\end{align}
where $\theta_i$ is the AoA of $i-$th arriving signal and $\tau_i$ is the corresponding ToA. Each 2-tuple member of the set $\mathcal{M}_e$ corresponds to a specific received signal coming from LoS, or a reflector. We assume that the BS can measure the variance set, $\mathrm{var}(\mathcal{M}_{e})$, defined as:
\begin{align}
\mathrm{var}(\mathcal{M}_{e})&=\{M_{{\bar \sigma}^2_{i}}\}, 1\leq i \leq \vert{\mathcal{M}_e}\vert \nonumber\\
&=\Big\{ \left({\sigma^2_{\theta_1},\sigma^2_{\tau_1}}\right),\dots,\left({\sigma^2_{\theta_{\vert{\mathcal{M}_e}\vert}},\sigma^2_{\tau_{\vert{\mathcal{M}_e}\vert}}}\right)\Big\},
\end{align}
where $\sigma^2_{\theta_i}$ and $\sigma^2_{\tau_i}$ represent the measured variance of AoA and ToA for the $i-$th received signal path. The values  $\sigma^2_{\theta_i}$ and $\sigma^2_{\tau_i}$ depend on the method of measurement, array structure, processing time and the level of noise and interference of the environment that are not in the scope of this paper.   
%\begin{align}
%f_{\theta_i}(\theta)&=\frac{1}{\sqrt{2 \pi \sigma^2_{\theta_i}}} e^{-\frac{(\theta-\theta_i)^2}{2 \sigma^2_{\theta_i}}}\\
%f_{\tau_i}(\tau)&=\frac{1}{\sqrt{2 \pi \sigma^2_{\tau_i}}} e^{-\frac{(\tau-\tau_i)^2}{2 \sigma^2_{\tau_i}}}
%\end{align}
Our goal is to design a function $\mathcal{G}(.)$ to localize the user based on the measurement sets, $\mathcal{M}_e$ and $\mathrm{var}(\mathcal{M}_{e})$:
\begin{equation}
\underset{{\mathcal{G}(.)}}{\mathrm{argmin}}{~\mathbb{E}\Big\{{\|\mathcal{G}\left({{\mathcal{M}_e},\mathrm{var}(\mathcal{M}_{e}) \vert \mathcal{M}_s}\right)-{\bf p}_u\|_2 \Big\}}}.
\end{equation}
Our technique involves two main steps. The first objective is to develop an approach to measure the effective parameters of the environment $\mathcal{M}_s$. Next, we aim to determine the localizing function $\mathcal{G}(.)$.
\subsection{Order of Reflections}
The order of reflection refers to the number of times a wave or signal undergoes reflection off surfaces within an environment while travelling between a transmitter and a receiver. Zero-order reflection represents the line of sight (LoS), with higher orders indicating paths involving multiple reflections. Typically, paths with more than one order of reflection exhibit weaker effects compared to the zero-order and the first-order reflection paths. This is due to their longer propagation paths and relatively small reflection coefficients. Consequently, we exclude the impact of reflectors beyond the first order, as their effects are considered negligible in our analysis. Furthermore, these effects can be mitigated by filtering paths with delays exceeding a specific value. Therefore, our focus is primarily on the zero-order and the first-order reflection paths \cite{YangJie,MendrzikRico}.
\subsection{Finding Physical Locations of Environmental Reflectors}
 For any $(\theta_i,\tau_i) \in \mathcal{M}_e$, the locus of points satisfying $\tau_i$ is an ellipse. This ellipse can be modeled as a set of points ${\bf p} \in \mathcal{PE}_i$, where the sum of the distances $\vert{{\bf p}-{\bf p}_{u}}\vert$ and $\vert{{\bf p}-{\bf p}_{b}}\vert$ has the constant value $c_0 \tau_i$ where $c_0$ and $\tau_i$ represent the speed of light and the ToA for the $i$-th measurement in the set $\mathcal{M}_e$, respectively. In other words, we have:
 \begin{equation}
 \label{ellipsequation}
 \mathcal{PE}_i=\bigl\{{{\bf p} \in \mathbb{R}^2 \vert~ \vert{{\bf p}-{\bf p}_{u}}\vert + \vert{{\bf p}-{\bf p}_{b}}\vert =c_0 \tau_i }\bigr\}.
 \end{equation}
%where $\vert{{\bf p}-{\bf p}_{u}}\vert$ represents magnitude of the vector ${{\bf p}-{\bf p}_{u}}$.
Similarly, each angle  $\theta_i$ in $\mathcal{M}_e$ represents a line which can be defined as the set of points as follows:
\begin{equation}
\label{lineequation}
 \mathcal{PL}_i=\biggl\{{{\bf p}=(x,y) \in \mathbb{R}^2 \Big{\vert}~ \frac{y-y_b}{x-x_b} = \tan{(\theta_i)} }\biggr\}.
 \end{equation} 
 We can find two different points by intersecting $\mathcal{PE}_i$ and $\mathcal{PL}_i$ (${\bf p}_{s_i}=\mathcal{PE}_i \cap \mathcal{PL}_i$), but only one of them is acceptable, where ${\bf p}_{b}-{\bf p}_{s_i}$ has the same direction as the AoA to the BS. Considering the location of the \color{black} receiver \color{black} ${\bf p}_b=(0,0)$, the location of the reflector from the measured parameters $\left({\tau_i, \theta_i}\right)$ can be represented as:
 \begin{equation}
 {\bf p}_{s_i} = (x_{s_i},y_{s_i}) = \mathcal{J}\left({(\tau_i,\theta_i), {\bf p}_u }\right).
 \end{equation} 
\begin{lemma}
{\it Considering ${\bf p}_b=(0,0)$ and $c_0 \tau_i > \lvert{{\bf p}_u }\rvert$ the location of the corresponding reflector ${\bf p}_{s_i}=(x_{s_i},y_{s_i})$ can be calculated as follows:
 \begin{align}
 \label{location1}
 x_{s_i} &= \frac{(c^2_0 \tau^2_i - \lvert{{\bf p}_u }\rvert^2) \cos(\theta_i)}{2 (c_0 \tau_i - \lvert{{\bf p}_u}\rvert \cos(\theta_i-\phi))}\\
 \label{location2}
 y_{s_i} &=x_{s_i} \tan{(\theta_i)},
 \end{align}
where $\phi$ represents the angle of the line connecting the transmitter to the receiver with respect to the x-axis. %Therefore, we have $\phi = \tan^{-1}{\left({{y_u}/{x_u}}\right)}$.}
}
\end{lemma} 
\begin{IEEEproof}
The proof is presented in Appendix \ref{appendix1}. 
\end{IEEEproof}
When the measurements are noisy, the location of each reflector  $s_i$ can be modeled by the mean  $\mu_{s_i} = \left({x_{s_i},y_{s_i}}\right)$ and variance  ${\bf{V}}_i$ which has been calculated in Appendix \ref{appendix1}. In the next section, we prepare an efficient strategy for user placement to gather sufficient information from the environment to construct the reflection map.
\section{User Placement Strategy for the Reflection Map Construction}
\label{reconstructionstage}
Recognizing that the composition of the set $\mathcal{M}_s$ plays a crucial role in the localization procedure, we introduce an approach to identifying the positions of effective reflectors in the environment. Before delving into the details, we establish a lemma that provides a sufficient condition for perfect measurement of the set $\mathcal{M}_s$. First, we provide the following definitions.
\begin{figure}
  \centering
  \subfloat a){{\includegraphics[width=6cm]{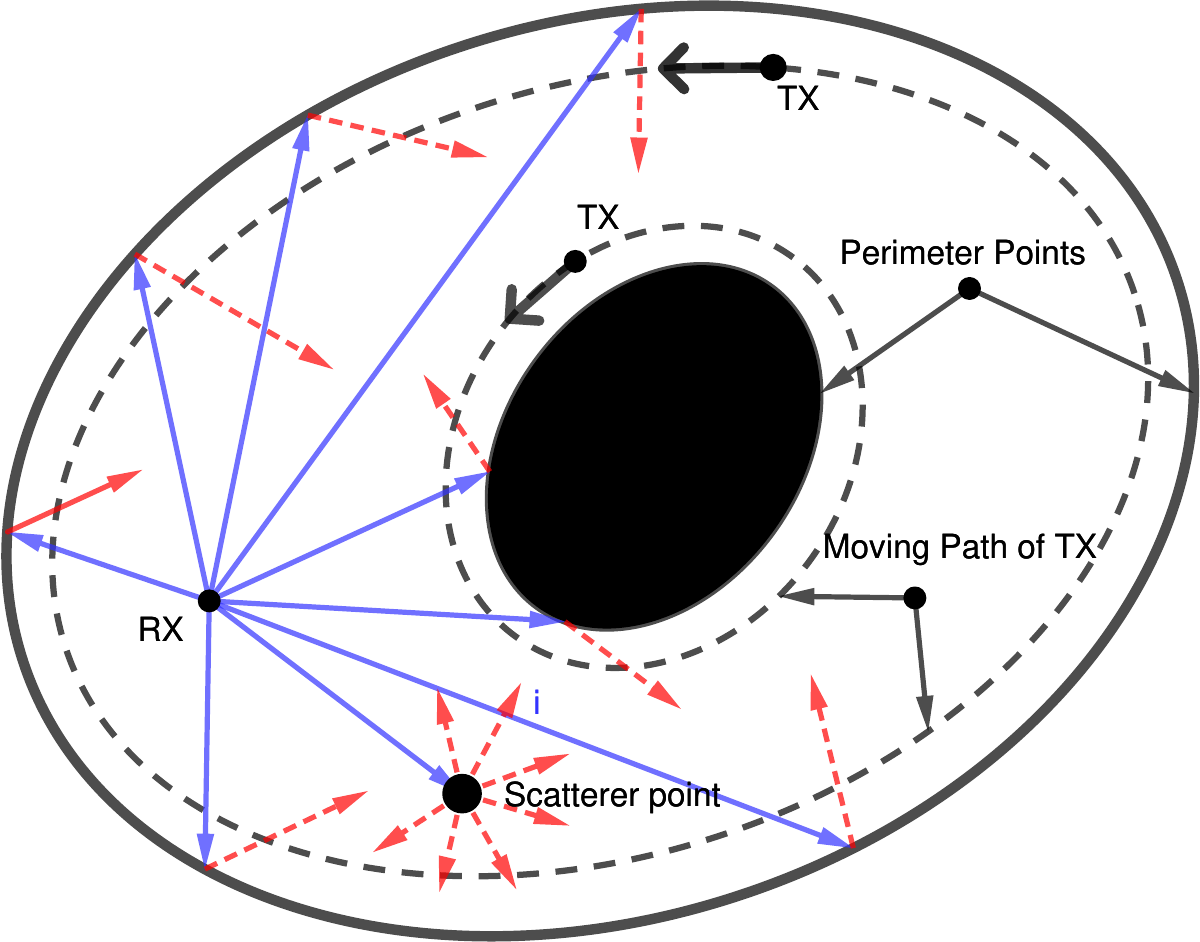} }}%
  \qquad
  \subfloat b){{\includegraphics[width=6cm]{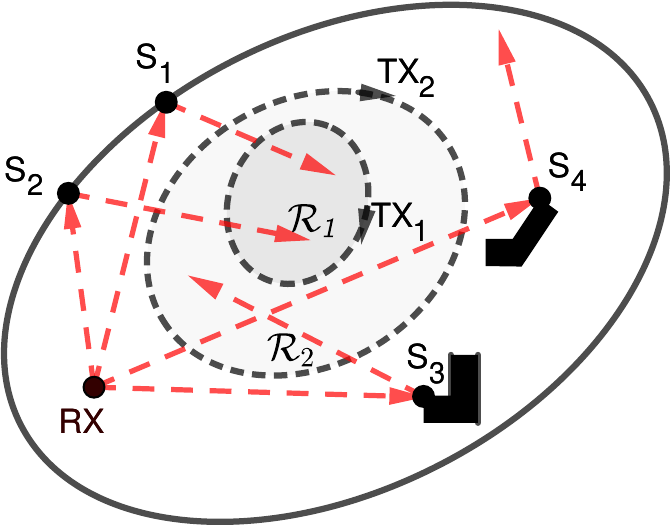} }}%
  \caption{Part (a) illustrates a confined boundary environment, featuring a transmitter with two trajectory paths and a receiver located in a specific position. The transmitter can vary its location for signal transmission. For each transmitter location, a subset of reflector objects situated along the boundary of the environment becomes activated. In (b), the transmitter, when positioned along the boundary of the empty localization regions  $\mathcal{R}_1$ and $\mathcal{R}_2$ can identify all the effective reflectors on these regions.}
  \label{ReflectionEnv}
\end{figure}
\begin{definition}
``{\it{Region of localization (RoL)}}'' refers to the area where the localization process is carried out.  
\end{definition}
For example, in Fig. \ref{ReflectionEnv} (a), the RoL has been bounded by a closed black line and a filled black elliptical shape.
\color{black}
\begin{definition}
``{\it{Effective reflector point}}'' is a point that can reflect the signal from a transmitter in a specific location in the {\it{RoL}}, to the receiver. 
\end{definition}
\color{black}
In Fig. \ref{ReflectionEnv} (b), two \color{black}RoLs \color{black}have been indicated by $\mathcal{R}_1$ and $\mathcal{R}_2$. The sets $\{s_1,s_2\}$ and $\{s_1,s_2,s_3\}$ are the effective reflectors of the localization regions  $\mathcal{R}_1$ and $\mathcal{R}_2$, respectively.
\begin{definition}
``{\it Blind spots}'' refers to those points inside the RoL. If a transmitter is placed at these points, there is no radio link with zero-order or first-order reflection paths between them and the receiver. 
\end{definition}
\color{black}
\begin{definition}
``{\it Test point}'' is a point within or on the boundary of an RoL where a transmitter is placed, and a set of measurements is taken at the receiver ($\mathcal{M}_e$). Each member of the set $\mathcal{M}_e$ corresponds to a LoS or a reflector with a specific location that can be calculated from \eqref{location1} and \eqref{location2}.
\end{definition}
\color{black}
\begin{lemma} 
\label{lemma2}
{\it The sufficient condition for identifying all effective reflectors on an empty RoL is met by moving a transmitter on the boundary of the RoL and measuring both AoA and ToA at the receiver.}
\end{lemma} 
\begin{IEEEproof}
 As depicted in Fig. \ref{ReflectionEnv} (b), if a transmitter within an RoL establishes first-order reflection paths (not in the blind spots) with the receiver, it activates reflection points outside the RoL in the set  $\mathcal{M}_s$. A line connecting the location of these reflection points to the location of the transmitter intercepts the boundary of the RoL at least at one point. 

In Fig. \ref{ReflectionEnv} (b), the trajectory of \color{black}a \color{black} transmitter (depicted by dashed lines) remaining on the boundary of the RoL (e.g., $\mathcal{R}_1$ and $\mathcal{R}_2$) enables it to intercept all the reflection rays from effective reflectors, indicated by the red dashed vectors. Consequently, when the transmitter emits a signal and move on the boundary of RoL, it triggers all the corresponding intercepted lines in the opposite direction. Therefore, by measuring both the AoA and ToA at the receiver and referring to \eqref{location1}, \eqref{location2} the receiver can accurately determine the location of the effective reflector point responsible for generating the intercepted ray. By moving a transmitter along the boundary of each RoL, we ensure knowledge of the locations of all effective reflector points corresponding to the RoL, thereby completing the proof of this lemma.
\end{IEEEproof}
\begin{lemma}
\label{subregionsreflectors}
{\it For any two empty localization regions, denoted as $\mathcal{R}_1$ and $\mathcal{R}_2$, with their corresponding sets of effective reflectors, denoted as $\mathcal{M}^{({\mathcal{R}_1})}_s$ and $\mathcal{M}^{({\mathcal{R}_2})}_s$, respectively, if $\mathcal{R}_1 \subseteq \mathcal{R}_2$, then $\mathcal{M}^{({\mathcal{R}_1})}_s \subseteq \mathcal{M}^{({\mathcal{R}_2})}_s$.}
\end{lemma}
\begin{IEEEproof}
Since any rays traversing through $\mathcal{R}_1$ must intersect with the boundary of the larger region, $\mathcal{R}_2$, every effective reflector point for the region $\mathcal{R}_1$ is an effective reflector for the region $\mathcal{R}_2$. Thus, $\mathcal{M}^{({\mathcal{R}_1})}_s \subseteq \mathcal{M}^{({\mathcal{R}_2})}_s$.
\end{IEEEproof}
\color{black}
In Fig. \ref{ReflectionEnv} (b), the transmitter, positioned along the boundary of the unoccupied region within $\mathcal{R}_1$, can effectively identify a subset of all reflector points which is represented by $\mathcal{M}^{({\mathcal{R}_1})}_s$. Specifically, this pertains to reflector points associated with ${s_1, s_2}$. Extending this scenario to $\mathcal{R}_2$, where $\mathcal{R}_1 \subseteq \mathcal{R}_2$, allows us to pinpoint reflector points for ${s_1, s_2, s_3}$ within $\mathcal{M}^{({\mathcal{R}_2})}_s$. 
\begin{lemma}
A sufficient condition to identify all the effective reflector points in an environment is to move a TX sufficiently close to its boundaries with a distance of $\zeta$, where $\zeta \rightarrow 0^+$.
\end{lemma}
\begin{IEEEproof}
Let us denote the path of moving a TX close to its boundary by $\mathcal{C}_{\zeta},~\zeta \rightarrow 0^+$. It is evident that every close RoL can be divided into many empty RoLs. Additionally, every empty RoL inside the environment is bounded by the physical boundary of the environment ($\mathcal{C}_{\zeta}$) and the boundary of infinite space ($\mathcal{C}_{\infty}$). All the RoLs are subsets of the RoL surrounded by $\mathcal{C}_{\zeta}$ and $\mathcal{C}_{\infty}$. Therefore, every effective reflector on every RoL can be identified by moving on the boundary of $\mathcal{C}_{\zeta}$, where $\zeta \rightarrow 0^+$, and $\mathcal{C}_{\infty}$. Since every NLoS propagation path intercepting $\mathcal{C}_{\infty}$ also intercepts $\mathcal{C}_{\zeta}$, it is sufficient to move the TX along $\mathcal{C}_{\zeta}$ to identify all the effective reflector points in the environment, thus proving the claim of this lemma.
\end{IEEEproof}
Notably, in Fig. \ref{ReflectionEnv} (a), the trajectory of the TX (illustrated by dashed lines) remaining proximate to the environment’s boundary enables it to intercept all reflection rays within the environment, as indicated by the red vectors. Consequently, when the TX emits a signal, it triggers the corresponding intercepted ray in the opposing direction. Measuring both the AoA and ToA at the RX, and referring to equations \eqref{location1} and \eqref{location2}, the receiver can accurately determine the location of the reflector point responsible for generating the intercepted ray. This approach successfully identifies all the effective reflectors. 
\color{black}
\begin{figure}[!t]
  \centering  \includegraphics[width=0.4\textwidth]{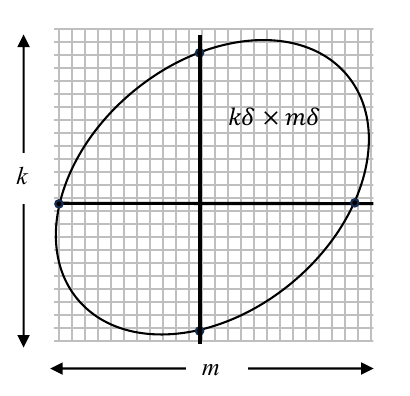}
  \caption{A convex RoL divided by $m$ vertical and $k$ horizontal lines, where each parallel line spaced by value  $\delta$. The length of the side of RoL is at least $2~\max(m,n) \delta$.}
  \label{Info2}
\end{figure}

\begin{lemma}
{\it In a two-dimensional application, the necessary number of test points for identifying effective reflector points of a convex RoL of $\mathcal{R}$ is bounded by $ n' \leq O(2\gamma \sqrt{n})$, where $n \rightarrow \infty$ represents the number of test point required to reconstruct the radio map of $\mathcal{R}$, and $2 \sqrt{\pi} \leq \gamma$.}
\end{lemma}

\begin{IEEEproof}
Referring to Lemma \ref{lemma2}, within an RoL, identifying all effective reflectors is achieved by moving a transmitter along the boundary of the region and measuring both AoA and ToA at the receiver. Thus, as the number of test points on the boundary of the localization region tends to infinity, it guarantees the identification of all effective reflector objects.

Consider a localization region $\mathcal{R}$ divided by $m$ vertical and $k$ horizontal lines, with each parallel line spaced by $\delta$ units. \color{black} These horizontal and vertical lines generate many partitions with the area of $\delta^2$. \color{black} For radio map reconstruction, a transmitter is placed in the middle of all partitions in the region  $\mathcal{R}$, and signal characteristics are measured at the receiver. As $m$ and $k$ approach infinity and subsequently $\delta$ approaches 0, the radio map for this region can be precisely reconstructed, with the number of test points being $n=\frac{{S}_{A}}{\delta^2}$, where ${S}_{A}$ represents the area of the region $\mathcal{R}$.

\color{black} To reconstruct the reflection points, we can position a TX at the intersection points of the horizontal and vertical lines with the boundary of the RoL. As the spacing parameter $\delta$ approaches 0 and both $m$ and $k$ approach infinity, the number of intersection points tends to infinity.
 \color{black} According to the conclusion of Lemma 2, when the number of test points on the boundary of the localization region reaches infinity, the complete effective reflector map can be reconstructed. \color{black} The number of test points on the boundary of the region is denoted as $n' = 2(m+k)$, representing the number of intersection points of horizontal and vertical lines with the boundary of the region. \color{black} Additionally, the length of the boundary of the RoL can be lower bounded by $2\delta~\mathrm{max}(m,k) \leq l_\mathcal{R}$. \color{black}The inequality arises from the fact that the horizontal or vertical lines have at least $2~ \mathrm{max}(m,k)$ intersection points with the boundary of the region, where the length of the boundary between each neighbouring horizontal or vertical line is at least $\delta$. \color{black} Since $m+k < 2~\mathrm{max}(m,k)$, from $2\delta~\mathrm{max}(m,k) \leq l_\mathcal{R}$, we can conclude that $n'=2(m+k) \leq \frac{2 l_\mathcal{R}}{\delta}$. %Finally, it follows that $ n'=2~(m+k) \leq \frac{2 l_\mathcal{R}}{ \delta}$.

Consequently, the number of test points required to identify effective reflector points within this region is directly upper-bounded proportionally by the side length of $\mathcal{R}$ ($ n' \leq \frac{2 l_{\mathcal{R}}}{\delta}$), whereas for conventional radio map identification methods, the number of test points is proportional to the area of the localization region ($n = \frac{{S}_{A}}{\delta^2}$). Therefore, it can be concluded that $n' \leq O\left(2 \gamma \sqrt{n} \right)$, where $\gamma=\frac{l_{\mathcal{R}}}{\sqrt{{S}_{A}}}$. The value of $\gamma$ depends on the geometric shape; for example, for a circle, it is $2\sqrt{\pi}$, and for a square, it is $4$. The minimum value  $\gamma$ is for a circular shape, where $\frac{l_\mathcal{R}}{\sqrt{{S}_{A}}} = 2\sqrt{\pi}$. Therefore, in a two-dimensional application, the necessary data size for identifying an environment using the reflector map is upper bounded by $ n' \leq O(2\gamma \sqrt{n})$, where $n$ represents the number of test points needed to reconstruct the radio map.
\end{IEEEproof}
\color{black} In the next section we show that how we can use the data from the test points to construct effective reflector points map. \color{black}
\section{Effective Reflector Points Construction}
Consider a scenario with a single reflector point located at ${\bf p}_s$ in an environment containing a transmitter and a receiver. In each transmission time duration \(T_s\), the receiver can measure the AoA and ToA within a certain margin of error from their true values. Utilizing these parameters, the receiver can estimate the location of the reflector point, denoted as \(\hat{\bf p}_s\). However, \(\hat{\bf p}_s\) is a random variable with a mean of \({\bf p}_s\) and a specific distribution due to measurement inaccuracies.

Since it's impractical to precisely pinpoint the location of reflectors, we define an area around each measured reflector point where the reflector is likely to fall with a probability greater than \(1-\epsilon\). Therefore, we define the concept of the {\it ``Covering Sheaf''} of the set \(\mathcal{M}_s\) as follows:

\begin{definition}
The{\it ``Covering Sheaf''} \(\mathcal{S}_{\epsilon}(\mathcal{M}_s)\) is a continuous bounded set of points in the x-y plane with the smallest volume of \(\mathrm{Vol}\left({\mathcal{S}_{\epsilon}(\mathcal{M}_s)}\right)\) such that:
\begin{equation}
\mathbb{P}\left({{\bf p} \in \mathcal{S}_{\epsilon}(\mathcal{M}_s)}\right) \geq 1-\epsilon, \quad {\bf p} \in \mathcal{M}_s.
\end{equation}
\end{definition}
\color{black} 

The set $\mathcal{S}_{\epsilon}(\mathcal{M}_s)$ is defined based on the premise that during the measurement of the set $\mathcal{M}_s$, only deflected versions of its members are accessible. By defining $\mathcal{S}_{\epsilon}(\mathcal{M}_s)$, we ensure that all points in the set $\mathcal{M}_s$ are members of $\mathcal{S}_{\epsilon}(\mathcal{M}_s)$ with a probability of at least $1-\epsilon$, where $\epsilon$ is chosen to be sufficiently small. For $\epsilon \rightarrow 0$, the covering sheaf of $\mathcal{S}_{\epsilon}(\mathcal{M}_s)$ is large enough to contain all the possible reflectors with the highest probability, while for $\epsilon \rightarrow 1$ the volume of it converges to zero. In this paper, our focus is on the case where $\epsilon \rightarrow 0$, which indicates that we have access to the region of reflectors almost surely. Henceforth in the paper, when referring to $\mathcal{M}_s$ or any related set as a subset, it specifically denotes $\mathcal{S}_{\epsilon}(\mathcal{M}_s)$ or a subset of it. For ease of notation, we will continue to use $\mathcal{M}_s$ or similar notation such as $\mathcal{M}^{(\mathcal{R})}_s$ to denote subsets of $\mathcal{S}_{\epsilon}(\mathcal{M}_s)$.
\color{black}

Lemma \ref{lemma2} provides a sufficient condition for identifying all the effective reflector objects in a specific region. In this section, we present a criterion for determining the required number of test points on the boundary of a region $\mathcal{R}$. By altering the transmitter's location on the boundary $\mathcal{R}$, we can gather information from the effective reflector points that have an impact on the region $\mathcal{R}$. This allows us to identify the set of effective reflectors $\mathcal{M}^{(\mathcal{R})}_{s} \subseteq \mathcal{M}_s$.
\begin{definition}
The discrete representation of the continuous set $\mathcal{M}^{(\mathcal{R})}_s$ is denoted by the set $\mathcal{M}^{(\mathcal{R})}_d$. This representation is obtained by locating a TX along the boundary of $\mathcal{R}$ at separated points and measuring ToA and AoA at RX. The resulting measurements are then mapped to their locations which represent the set $\mathcal{M}^{(\mathcal{R})}_d$.  
\end{definition}

 The set $\mathcal{M}^{(\mathcal{R})}_s$ can be represented by the two dimensional function $M(x,y)$, where we can define it as:
\begin{equation}
\label{Mxy}
M(x,y)=\left\{ 
  \begin{array}{ c l }
    1 & \quad \textrm{if } (x,y) \in \mathcal{M}^{(\mathcal{R})}_s \\
    0                 & \quad \textrm{if } (x,y) \notin \mathcal{M}^{(\mathcal{R})}_s
  \end{array}.
\right.
\end{equation}
The function $M(x,y)$ is an indicator function, classifying points in space as either belonging to the set of effective reflector points, represented by the value 1, or outside this set, represented by the value 0.
The 2-dimensional Fourier transform of the above continuous function can be represented by $M_{\lambda}(\lambda_1,\lambda_2)=\mathcal{F}\{{M(x,y)}\}$ as follows:
\begin{equation}
\label{2dF}
M_{\lambda}(\lambda_1,\lambda_2)={\iint_{-\infty}^{+\infty}{M(x,y)e^{-2\pi j(\lambda_1 x+\lambda_2 y)}dxdy}}.
\end{equation}
 The parameters $\lambda_1$ and $\lambda_2$ are spatial frequency. In the above relation we can assume that for $\lvert \lambda_1 \rvert> \lambda_{\mathrm{m}}, \lvert \lambda_2 \rvert > \lambda_{\mathrm{m}}$, the value  $M_{\lambda}(\lambda_1,\lambda_2) \approx 0$. This comes from the fact that $M(x,y)$ in its domain is an energy-limited function as follows:
 \begin{equation}
 \iint_{x,y \in \mathcal{M}^{(\mathcal{R})}_s}{\lvert M(x,y) \rvert^2 dx dy} < \infty.
 \end{equation}
 From \eqref{Mxy} we have:
\begin{equation}
\label{intf}
M_{\lambda}(\lambda_1,\lambda_2)={\iint_{\mathcal{M}^{(\mathcal{R})}_s}{e^{-2\pi j(\lambda_1 x+\lambda_2 y)}dxdy}}.
\end{equation}
\color{black} The above relation comes from the fact that for $(x,y) \notin \mathcal{M}^{(\mathcal{R})}_s$ the value  $M(x,y)$ is zero therefore, the integral can be done on those points which are in the bounded area of $\mathcal{M}^{(\mathcal{R})}_s$. The above relation can be estimated by the following relation:\color{black}
\begin{equation}
%\label{disf}
{M}_{\lambda}(\lambda_1,\lambda_2)=\lim_{L \rightarrow \infty}\frac{\mathrm{Vol}(\mathcal{M}^{(\mathcal{R})}_s)}{L}\sum_{i=1}^{L}{M(x_i,y_i) e^{-2\pi j(\lambda_1 x_i+\lambda_2 y_i)}},
\end{equation}
where in the above relation, $(x_i, y_i)$ are distributed across the area of $\mathcal{M}^{(\mathcal{R})}_s$ at equal distances. The given relation represents the summation form of the integral relation in equation \eqref{intf}, where $dxdy$ has been substituted by $\frac{\mathrm{Vol}(\mathcal{M}^{(\mathcal{R})}_s)}{L}$. 

By randomly placing the user multiple times at the boundary of $\mathcal{R}$ and obtaining measurements at the BS, we can acquire numerous samples  $(x_i, y_i),~1\leq i \leq N$, corresponding to the locations of the reflector points, depending on the transmitter's position. Therefore, the output of our test points on the boundary of the localization region $\mathcal{R}$ is a set denoted as $\mathcal{M}^{(\mathcal{R})}_{d}$, which is a subset of the set of effective reflectors $\mathcal{M}^{(\mathcal{R})}_{s}$. We define the discrete sampled version of $M(x,y)$ as follows:
\begin{equation}
M_d(x,y)=\sum_{i=1}^N{M(x_i,y_i) \delta(x-x_i,y-y_i)},
\end{equation}
where $(x_i,y_i) \in \mathcal{M}^{(\mathcal{R})}_d \subseteq \mathcal{M}^{(\mathcal{R})}_s$. These samples have a specific distribution across the area of the set $\mathcal{M}^{(\mathcal{R})}_s$. For now, we will consider it to be a uniform distribution with the probability density function (PDF) $f_R({\bf p})$, ${\bf p} = (x, y)$, as follows:
\begin{equation}
f_R({\bf p}) = \left\{
  \begin{array}{cl}
    \frac{1}{\mathrm{Vol}\left(\mathcal{M}^{(\mathcal{R})}_s\right)} & \quad \textrm{if } {\bf p} \in \mathcal{M}^{(\mathcal{R})}_s, \\
    0 & \quad \textrm{if } {\bf p} \notin \mathcal{M}^{(\mathcal{R})}_s.
  \end{array}
\right.
\end{equation}
\begin{lemma}
The following estimator is an unbiased and asymptotically consistent estimator of ${M}_{\lambda}(\lambda_1,\lambda_2)$ as $N$ increases to infinity.
\begin{equation}
\label{disf}
\tilde{M}_{\lambda}(\lambda_1,\lambda_2)=\frac{\mathrm{Vol}\left({\mathcal{M}^{(\mathcal{R})}_s}\right)}{N}\sum_{i=1}^{N}{ e^{-2\pi j(\lambda_1 x_i+\lambda_2 y_i)}}.
\end{equation} 
\end{lemma}
\begin{IEEEproof}
The proof has been given in Appendix \ref{proofunbiased}. 
\end{IEEEproof}
 In general, when we perform non-uniform sampling from the set $\mathcal{M}^{(\mathcal{R})}_s$ with a general distribution  $f_R({\bf p})$, this approach leads us to the Fourier transform of the following function:
\begin{equation}
M(x,y)f_R({\bf p})=\left\{ 
  \begin{array}{ c l }
    f_R({\bf p}) & \quad \textrm{if } {\bf p}=(x,y) \in \mathcal{M}^{(\mathcal{R})}_s \\
    0                 & \quad \textrm{if } {\bf p}=(x,y) \notin \mathcal{M}^{(\mathcal{R})}_s
  \end{array}
\right.
\end{equation}
It has been proven in Appendix \ref{FourierRepresentation} that the following estimator can attain the Fourier transform  $M(x,y)f_R({\bf p})$ with zero mean and a variance that approaches zero as $N$ goes to infinity.
\begin{equation}
\label{disf3}
\tilde{M}_{\lambda}(\lambda_1,\lambda_2)=\frac{1}{N}\sum_{i=1}^{N}{ e^{-2\pi j(\lambda_1 x_i+\lambda_2 y_i)}}.
\end{equation}
A question that may arise here is whether the value of the function $M(x,y)f_R({\bf p})$ matches that of $M(x,y)$. However, in the localization section, we will demonstrate that all the information we need is contained within the domain of $M(x,y)$, where both $M(x,y)$ and $M(x,y)f_R({\bf p})$ have the same domain. In Appendix \ref{recoverysetMsfromMd}, we demonstrate how we can reconstruct the set of reflector points, denoted as $\mathcal{M}^{(\mathcal{R})}_{s}$, from the set $\mathcal{M}^{(\mathcal{R})}_{d}$. In the next section, we propose a lower bound for localization accuracy for the case where $\mathrm{SNR}$ goes to infinity. Based on the constructed set $\mathcal{M}^{(\mathcal{R})}_{s}$, we propose an algorithm for localizing a user with the measurement set $\mathcal{M}_e$.
\section{Localization Accuracy Lower Bound and Proposed Strategy}
\subsection{Localization Accuracy Lower Bound}
In this section, first, we focus on finding a lower bound on the localization accuracy when $\mathrm{SNR}$ goes into infinity then we outline our localization strategy.  To elucidate the impact of the set $\mathcal{S}_{\epsilon}(\mathcal{M}_s)$ on the accuracy of our localization strategy, we present the following theorem. However, before delving into this theorem, we introduce \color{black}three \color{black}definitions.
\begin{definition}
{\it ``Reflectivity parameter of an environment''} $n_r \in \{0,1,2,\dots \}$ is a number that shows the typical number of first-order reflection paths from the user to the BS.
\end{definition}

\color{black}In essence, $n_r$ serves as an indicator of the reflective nature of an environment. A larger value of $n_r$ corresponds to a more intricate environment characterized by a higher count of first-order reflectors. 
\begin{definition}
\label{defsu}
The \textit{Ambiguity set}, denoted by $\mathcal{S}_u$, arises from localization errors and can be defined as follows:
\small
\begin{equation}
\mathcal{S}_u = \left\lbrace \mathcal{T}\left( {\bar {\bf Y}}^{(i)}\, \vert \, \mathcal{S}_{\epsilon}\left({\mathcal{M}_s}\right), \mathcal{S}_{A} \right) - {\bf p}_u \mid i \in \{1, \dots, n\} \right\rbrace,
\end{equation}
\normalsize
where $n$ goes to infinity and $\hat{\bf p}_u^{(i)} = \mathcal{T}\left( {\bar {\bf Y}}^{(i)}\, \vert \, \mathcal{S}_{\epsilon}\left({\mathcal{M}_s}\right), \mathcal{S}_{A} \right)$ represents any localization function based on the $i$-th observation ${\bar {\bf Y}}^{(i)} \in \mathbb{C}^{J \times K}$ at RX and the given $\mathcal{S}_{\epsilon}\left({\mathcal{M}_s}\right)$. In this case $J$ and $K$ represent number of array antenna elements and the number of analyzed time snapshots, respectively. 
\end{definition}
\color{black}
\color{black}
\begin{definition}
The \textit{Ambiguity area}, denoted by $\mathrm{Vol}\left(\mathcal{S}_{\epsilon}\left(\mathcal{S}_u\right)\right)$, represents by the area of the covering sheaf of the set $\mathcal{S}_u$. For simplicity of notation we represents $\mathrm{Vol}\left(\mathcal{S}_{\epsilon}\left(\mathcal{S}_u\right)\right)$ with $\mathrm{Vol}\left(\mathcal{S}_u\right)$.
\end{definition}
\color{black}
\begin{theorem}
\label{thm1}
{\it {Given a uniform distribution of reflectors across the covering sheaf  $\mathcal{S}_{\epsilon}\left({\mathcal{M}_s}\right)$, for the RoL of $\mathcal{S}_{A}$ with the volume of $\mathrm{Vol}\left({\mathcal{S}_{A}}\right)$ and non-LoS links, as $\mathrm{SNR} \rightarrow \infty$, $\mathrm{Vol}(\mathcal{S}_u)$ is lower-bounded as follows:}}
\begin{equation}
\label{res1}
\mathrm{Vol}(\mathcal{S}_u) > \mathrm{Vol} \left({\mathcal{S}_{A}}\right) 2^{-n_r \log_{2}\left({1+\frac{ \mathrm{Vol}(\mathcal{S}_{A})}{\mathrm{Vol}\left({\mathcal{S}_{\epsilon}(\mathcal{M}_s)}\right)}}\right)}.
\end{equation}
\end{theorem}
\begin{IEEEproof}
See the details of the proof in Appendix \ref{appendix0}. 
\end{IEEEproof}
The above theorem reveals a significant relationship between $n_r$  and the lower bound of the ambiguity area. Essentially, these two parameters, $n_r$ and $\mathrm{Vol}\left({\mathcal{S}_{\epsilon}(\mathcal{M}_s)}\right)$, are intrinsically linked. A greater physical volume of reflectors \color{black} or equivalently $\mathrm{Vol}\left({\mathcal{S}_{\epsilon}(\mathcal{M}_s)}\right)$, \color{black} tends to result in a higher value of $n_r$, but it can decrease the value of the term $\log_{2}\left({1+\frac{ \mathrm{Vol}(\mathcal{S}_{A})}{\mathrm{Vol}\left({\mathcal{S}_{\epsilon}(\mathcal{M}_s)}\right)}}\right)$.

For $n_r=0$, we do not have any LoS or non-LoS links, \color{black} referring \eqref{res1}, we have $\mathrm{Vol}(\mathcal{S}_u) > \mathrm{Vol}(\mathcal{S}_{A})$, \color{black}which means that without any RF links, we cannot identify the location of the user. For $n_r=1$, one may raise the question that even in the case of $\mathrm{Vol}(\mathcal{S}_{\epsilon}(\mathcal{M}_s)) \rightarrow 0$ (point reflector), we cannot identify the location of the transmitter because we have ambiguity on the circumference of a circle. In this case, the ambiguity area is on the circumference of a circle with zero thickness, which indicates that $\mathrm{Vol}(\mathcal{S}_u)>0$. Of course, to identify the location of the user around a single point, we need to have $n_r > 2$.

\color{black} As our measurements in the first phase become more accurate, the measured reflection points deviate less from their original locations, resulting in a lower value for $\mathrm{Vol} \left({\mathcal{S}_\epsilon(\mathcal{M}_s)}\right)$. Consequently, this increases the term $\log_{2}\left({1+\frac{ \mathrm{Vol}(\mathcal{S}_{A})}{\mathrm{Vol}\left({\mathcal{S}_{\epsilon}(\mathcal{M}_s)}\right)}}\right)$, indicating higher localization accuracy.\color{black}

\begin{figure}[!t]
  \centering  \includegraphics[width=0.6\textwidth]{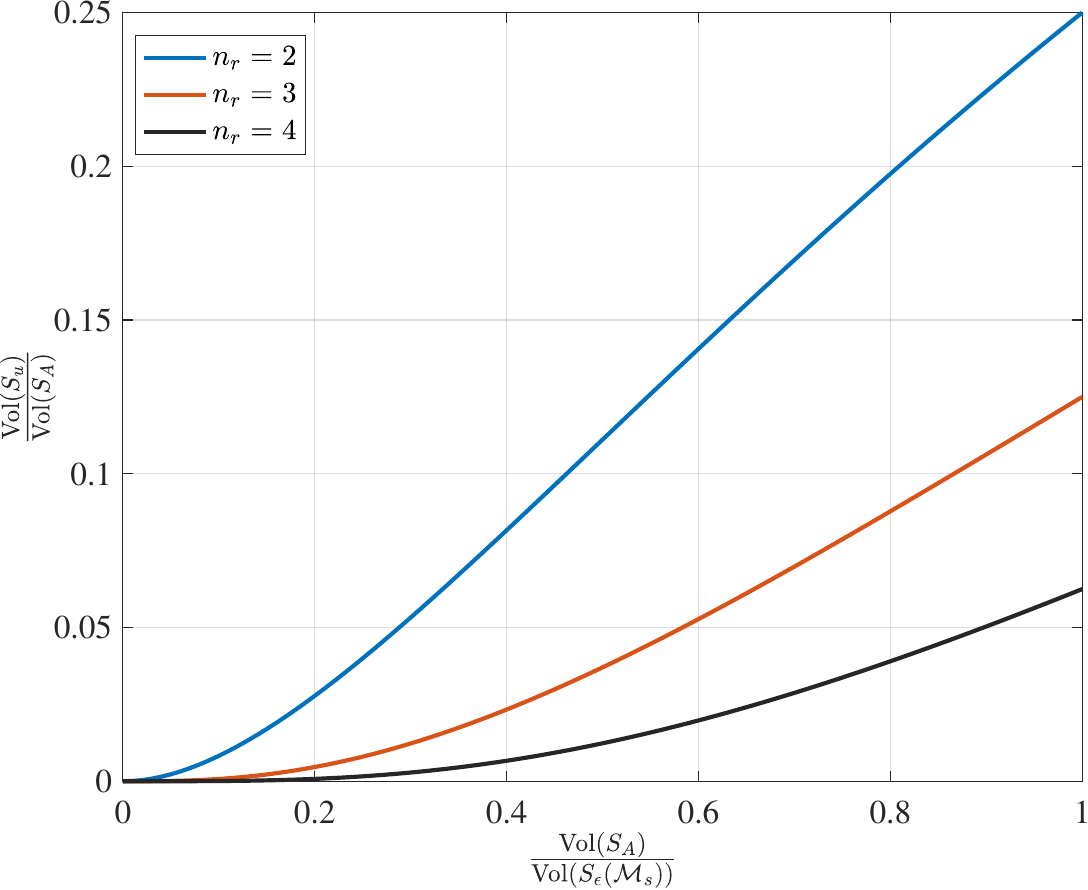}
  \caption{Lower bound  $\frac{\mathrm{Vol}(\mathcal{S}_u)}{\mathrm{Vol} \left({\mathcal{S}_{A}}\right)}$ as a function  $\frac{ \mathrm{Vol}(\mathcal{S}_{A})}{\mathrm{Vol}\left({\mathcal{S}_{\epsilon}(\mathcal{M}_s)}\right)}$ for different values  $n_r$.}
\label{lowerbound}
\end{figure}

Fig. \ref{lowerbound} represents the lower bound of $\frac{\mathrm{Vol}(\mathcal{S}_u)}{\mathrm{Vol} \left({\mathcal{S}_{A}}\right)}$ as a function of $\frac{ \mathrm{Vol}(\mathcal{S}_{A})}{\mathrm{Vol}\left({\mathcal{S}_{\epsilon}(\mathcal{M}_s)}\right)}$ for different values of $n_r$. 
\color{black} As the number of reflector points ($n_r$) increases, the value of the lower bound of $\mathrm{Vol}\left({\mathcal{S}_u}\right)$ decreases. This indicates that a higher number of reflectors leads to higher localization accuracy. Typically, increasing the number of reflectors increases the volume of the covering sheaf of reflector points. However, paradoxically, while increasing $n_r$ enhances localization accuracy, increasing the volume of the covering sheaf reduces localization accuracy. Therefore, adding artificial reflectors inside the environment with minimum volume and accurate locations can improve localization accuracy. For example, adding corner reflectors and reconfigurable intelligence surfaces (RIS) with exact locations and minimum physical volume can drastically improve the performance of localization. \color{black}
By strategically introducing such point reflectors, it becomes possible to enhance the value \color{black}of \color{black} $n_r$ and thus the environment's reflectivity, without substantially increasing the overall volume of reflectors. This strategy allows for the improvement of user location measurement accuracy in complex environments by increasing $n_r$ without increasing $\mathcal{S}_{\epsilon}(\mathcal{M}_s)$.
\begin{definition}
{\it Log-scale accuracy ratio} $R_a$, represents the accuracy of any localization procedure which can be defined by $R_a = \log_2{\frac{\mathrm{Vol}(\mathcal{S}_{A})}{\mathrm{Vol}(\mathcal{S}_u)}}$.
\end{definition}
From the above definition, it is obvious that a higher value of $R_a$ means higher localization accuracy and vice versa. From Theorem \ref{thm1}, we can find an upper-bound for $R_a$ as follows:
\begin{equation}
R_a < n_r \log_{2}\left({1+\frac{ \mathrm{Vol}(\mathcal{S}_{A})}{\mathrm{Vol}\left({\mathcal{S}_{\epsilon}(\mathcal{M}_s)}\right)}}\right).
\end{equation}

\subsection{Localization Strategy}
Consider we select a potential user location $\hat{\bf p}_u$. Leveraging the BS location, $\hat{\bf p}_u$ as the candidate location of the user, the measurement set $\mathcal{M}_e$ and $\mathrm{var}\left({\mathcal{M}_e}\right)$ at the BS, each 2-tuple $\left(\tau_i,\theta_i\right)$ is mapped to a specific location in the x-y plane, representing a single presumptive reflector point's location with a two-dimensional distribution and covariance matrix  ${\bf V}_i$. The subsequent step involves evaluating the alignment of these reflection points with the set $\mathcal{S}_{\epsilon}(\mathcal{M}_s)$ (where we use $\mathcal{M}_s$ interchangeably with the same concept). Ultimately, we determine a candidate location that exhibits the closest alignment probability with this set. In the preceding sections, we discussed strategies for environmental sensing, denoted by the set  $\mathcal{M}_s$. Now, let us consider a set of measured two-tuple members, $\mathcal{M}_e=\{(\tau_i,\theta_i), 1\leq i \leq n\}$, and their measurement variance set $\mathrm{var}\left({\mathcal{M}_e}\right)$. The BS scans the environment and measures the AoAs along with their corresponding ToAs and their variance set, and form $\mathcal{M}_e$ and $\mathrm{var}\left({\mathcal{M}_e}\right)$. For every value of the vector $\hat{\bf p}_u$, each 2-tuple member of the measurement set  $\mathcal{M}_e$ represents a reflection point which can be calculated by the function $\hat{\bf p}_{s_i} =\mathcal{J}\left({(\tau_i,\theta_i), \hat{\bf p}_u }\right)$ in \eqref{location1}, \color{black}\eqref{location2}\color{black}. \color{black} This function depends on the value  $\hat{\mathbf{p}}_u$ and maps each 2-tuple member of the measurement set $\mathcal{M}_e$ to a specific location that may lie in the set $\mathcal{M}_s$, out of $\mathcal{M}_s$ or $\varnothing$ where $\varnothing$ comes from the case where $\lvert {\hat{\mathbf{p}}_u-{\bf p}_b}\rvert > c_0 \tau_i$. \color{black}
We use the following maximization problem to estimate the location of the user:
 \begin{equation}
 \label{opt}
\arg \max_{\hat{\bf p}_u} { f_{P \vert \bar{M}}\left(\hat{\bf p}_u \vert \mathcal{M}_e\right)}.
 \end{equation}
The probability distribution function $f_{P \vert \bar{M}}\left(\hat{\bf p}_u \vert \mathcal{M}_e \right)$ can be represented as:
\small
\begin{equation}
f_{P \vert \bar{M}}\left(\hat{\bf p}_u \vert \mathcal{M}_e\right)=\frac{f_{\bar{M} \vert P}\left(\mathcal{M}_e \vert \hat{\bf p}_u\right) f_{P}\left({\hat{\bf p}_u}\right)}{f_{\bar{M}}\left({\mathcal{M}_e}\right)}.
\end{equation}
\normalsize
Since $f_P\left({\hat{\bf p}_u}\right)$ and $f_{\bar{M}}\left(\mathcal{M}_e\right)$ are the same for all possible locations  $\hat{\bf p}_u$, we can focus on the $f_{\bar{M} \vert P}\left(\mathcal{M}_e \vert \hat{\bf p}_u\right)$ which can be extended as follows:
\small
\begin{align}
f_{\bar{M} \vert P}&\left( \mathcal{M}_e \vert \hat{\bf p}_u\right)=f_{\bar{M} \vert P}\left( M_1,\dots,M_n \vert \hat{\bf p}_u\right)\\
&=\int_{{\bf s} \in \mathcal{M}_s}{f_{\bar{M} {S}\vert P}\left( M_1,\dots,M_n,{{\bf s}} \vert \hat{\bf p}_u\right)d{{\bf s}}}\\
&=\int_{\bar{\bf s} \in \mathcal{M}_s}{f_{ S}\left({{\bf s}}\right)f_{\bar{M} \vert P {S}}\left( M_1,\dots,M_n \vert \hat{\bf p}_u,{\bf s}\right)d{\bf s}}.
\end{align}
\normalsize
where $\mathcal{M}_e=\{{M_i},1\leq i \leq n\}$, ${\bf s} \in \mathcal{M}_s$. \color{black} In the above equations, the parameter $n$ represents the number of reflection points, typically denoted by $n_r$. \color{black} Considering ${\bf s}_j$ uniformly distributed across all the members of the set $\mathcal{M}_s$, the maximization of the above relation can be performed by maximizing:
\begin{equation}
\int_{{\bf s} \in \mathcal{M}_s}{f_{\bar{M} \vert P {S}}\left( M_1,\dots,M_n \vert \hat{\bf p}_u,{\bf s}\right)d{{\bf s}}}.
\end{equation}
For simplicity, we assume that all the measurements are independent from each other. Therefore, the above relation can be simplified as follows:
%\begin{figure*}[b]
%\hrule
%\footnotesize
\small
\begin{align}
\label{int1}
&\int_{{\bf s} \in \mathcal{M}_s}{f_{\bar{M} \vert P {S}}\left( M_1,\dots,M_n \vert \hat{\bf p}_u,{\bf s} \right)d {\bf s}}\nonumber\\
&=\int_{{\bf s} \in \mathcal{M}_s}{f_{M \vert P {S}}\left( M_1 \vert \hat{\bf p}_u,{\bf s}\right)\dots {f_{{{M} \vert P {S}}}\left( M_n \vert \hat{\bf p}_u,{\bf s}\right)d{\bf s}}}\\
%&=\int_{{\bf s}_n \in \mathcal{M}_s}\dots\int_{{\bf s}_1 \in \mathcal{M}_s}{\frac{{\mathrm{det}({\bf V}_{1})}^{-\frac{1}{2}}}{\sqrt{2 \pi}} e^{-\frac{1}{2}{\bf u}_{1j}{\bf V}^{-1}_{1}{\bf u}^T_{1j}}\dots \frac{{\mathrm{det}({\bf V}_{n})}^{-\frac{1}{2}}}{\sqrt{2 \pi}} e^{-\frac{1}{2}{\bf u}_{nj}{\bf V}^{-1}_{n}{\bf u}^T_{nj}} d{\bf s}_1\dots d{\bf s}_n}\\
&= \prod_{i=1}^{n}{\frac{{\mathrm{det}({\bf V}_{i})}^{-\frac{1}{2}}}{\sqrt{2 \pi}}}\int_{{\bf s} \in \mathcal{M}_s}{e^{-\frac{1}{2}{\bf u}_{1}{\bf V}^{-1}_{1}{\bf u}^T_{1}}\dots e^{-\frac{1}{2}{\bf u}_{n}{\bf V}^{-1}_{n}{\bf u}^T_{n}} d{\bf s}} \\
\label{int2}
&= \prod_{i=1}^{n}{\frac{{\mathrm{det}({\bf V}_{i})}^{-\frac{1}{2}}}{\sqrt{2 \pi}}}{\int_{{\bf s} \in \mathcal{M}_s}{e^{-\frac{1}{2} \sum_{i=1}^{n}{{\bf u}_{i}{\bf V}^{-1}_{i}{\bf u}^T_{i}}} d{\bf s}}},
\end{align}
\normalsize
%\end{figure*}
where ${\bf u}_{i}=\mathcal{J}\left({(\tau_i,\theta_i), \hat{\bf p}_u }\right)-{\bf s}$, and ${\bf V}^{-1}_{i}$ is the inverse of the covariance matrix from the $i-$th measurement member of the set $\mathcal{M}_e$, which depends on the received power and has been calculated in Appendix \ref{appendix1}. Therefore, the previous maximization problem can be replaced by maximizing the function
\begin{equation}
\label{scorefunction}
Q(\hat{\bf p}_u)= \int_{{\bf s} \in \mathcal{M}_s}{e^{-\frac{1}{2} \sum_{i=1}^{n}{{\bf u}_{i}{\bf V}^{-1}_{i}{\bf u}^T_{i}}} d{\bf s}}.
\end{equation}
 Therefore, we can focus on the problem  $\arg \max_{\hat{\bf p}_u} {Q(\hat{\bf p}_u)}$. As the computational of \eqref{scorefunction} is complex, in the next subsection we propose a simple pre-processing scheme that limits the search area.
\subsection{Pre-processing Technique}
In this subsection, we introduce a technique designed to select the region where the user is located with high probability. This approach reduces the area in which our searching algorithm operates, thus accelerating its computation. Our technique has the following steps:
\subsubsection{Dividing the Space}
Our technique begins by dividing the space into sectors defined by AoA. For each measured AoA, denoted as $\theta_i$ with $1 \leq i \leq n$, we create a specific angular range  $\pm \delta_i$, which we call it $\theta_i$ sector. We set $\delta_i$ to be sufficiently larger than $\sqrt{\mathrm{var}(\theta_i)}$, where we assume $\delta_i = k \sqrt{\mathrm{var}(\theta_i)}$ and $k>1$. The choice  $\delta_i$ ensures that the angular range around each $\theta_i$ is large enough to capture the variations in the AoA measurements, while not being overly broad.
\subsubsection{Sector-based Selection}
Within each $\theta_i$ sector, we select the points in the set  $\mathcal{M}_s$ that fall within the sector $\theta_i \pm \delta_i$, denoting this subset as $\mathcal{M}_s(\theta_i)$. By selecting points within the sector, $\mathcal{M}_s(\theta_i)$, we are already limiting the search space to locations that have AoA measurements consistent with $\theta_i$ from the reflection sets $\mathcal{M}_s$.
\subsubsection{Defining Local Regions}
We define the set $\mathcal{D}(\theta_i)$ as follows:
\small
\begin{equation}
\mathcal{K}(\theta_i) = \Big\{{\bf s}: c_0 \tau_0(\theta_i) <\lvert{{\bf s}-{\bf p}_{\theta_i}}\rvert+\lvert{{\bf p}_b-{\bf p}_{\theta_i}}\rvert < c_0 \tau_1(\theta_i)\Big\},
\end{equation}
\normalsize
where ${\bf p}_{\theta_i} \in \mathcal{M}_s(\theta_i)$. The values $\tau_0 (\theta_i)$ and $\tau_1 (\theta_i)$ are carefully chosen to cover the maximum and minimum delay values in the sector  $\theta_i$. This selection encompasses the points where the user is located with high probability.
\subsubsection{Final Localization Region}
The pre-localization region, denoted as $\mathcal{K}({\mathcal{M}_e})$, considers all individual regions $\mathcal{K}({\theta_i})$. It is calculated as the intersection of these regions as $\mathcal{K}({\mathcal{M}_e}) = \bigcap_{i=1}^{n}{\mathcal{K}({\theta_i})}$.

This technique provides a reliable method for determining the region most likely to contain the user's location based on the measured set  $\mathcal{M}_e$ and $\mathrm{var}\left({\mathcal{M}_e}\right)$.
\subsection{Gradient Ascent Algorithm for Optimization}
In \eqref{scorefunction}, the function $Q(\hat{\bf p}_u)$ is a multivariate function with a vector variable $\hat{\bf p}_u = (\hat{x}_u, \hat{y}u)$. To simplify the complexity of the searching algorithm, we can confine the limits of \eqref{scorefunction} to the set $\cup{j}\mathcal{M}_s \left({\theta_j}\right)$. Thus, for further simplification in the searching process, we can express \eqref{scorefunction2} as follows:

\begin{equation}
\label{scorefunction2}
Q(\hat{\bf p}_u)= \int_{{\bf s} \in \cup_{j}\mathcal{M}_s \left({\theta_j}\right)}{e^{-\frac{1}{2} \sum_{i=1}^{n}{{\bf u}_{i}{\bf V}^{-1}_{i}{\bf u}^T_{i}}} d{\bf s}}.
\end{equation}
Our objective is to employ an optimization procedure to find the highest peak within the region $\mathcal{K   mm}(\mathcal{M}_e)$. This is achieved through the following iterative steps:
\begin{enumerate}
    \item Start with an initial estimate $\hat{\bf p}_u^{(t)} \in \mathcal{K}({\mathcal{M}_e})$ at $t$.  
    \item Update the estimate using a gradient ascent step, governed by the equation:
    \begin{equation}
        \hat{\bf p}_u^{(t+1)} = \hat{\bf p}_u^{(t)} + \gamma \nabla Q\left(\hat{\bf p}_u^{(t)}\right).
    \end{equation}
    where $\gamma$ represents the step size, and $\nabla Q\left(\hat{\bf p}_u^{(t)}\right)$ denotes the gradient of the function $Q$ evaluated at $\hat{\bf p}_u^{(t)}$.
\end{enumerate}
In such cases, the gradient function can be approximated through numerical methods. One practical approach involves perturbing the components \(\hat{\mathbf{p}}_u^{(t)}\) along the \(x\) and \(y\) axes by small values \(\Delta x\) and \(\Delta y\). This numerical approach to estimating the gradient is commonly employed in situations where analytical solutions are challenging to derive. The algorithm is iterated until \(\lvert \hat{\mathbf{p}}_u^{(t)} - \hat{\mathbf{p}}_u^{(t-1)} \rvert\) is sufficiently small. To escape from local maxima, we repeat the procedure for a new set of starting points in the region \(\mathcal{K}(\theta_i)\) and select the point with the maximum value for the function \(Q(\cdot)\). In the next section, we analyze our simulation results on the reflection map construction, localization technique, and our lower bound on localization accuracy.              
\section{Simulation Results}
\color{black} In this section, we present simulation results where we consider a transmitter capable of sending a signal through an environment with many reflectors.
\begin{figure}
    \centering
    \subfloat a){{\includegraphics[width=6cm]{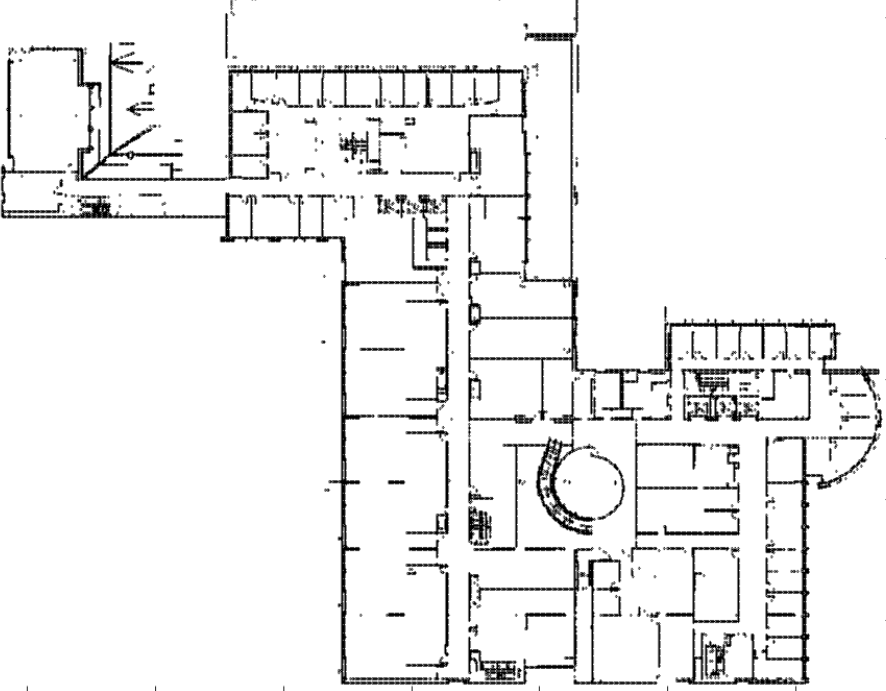}}}%
    \qquad
    \subfloat b){{\includegraphics[width=6cm]{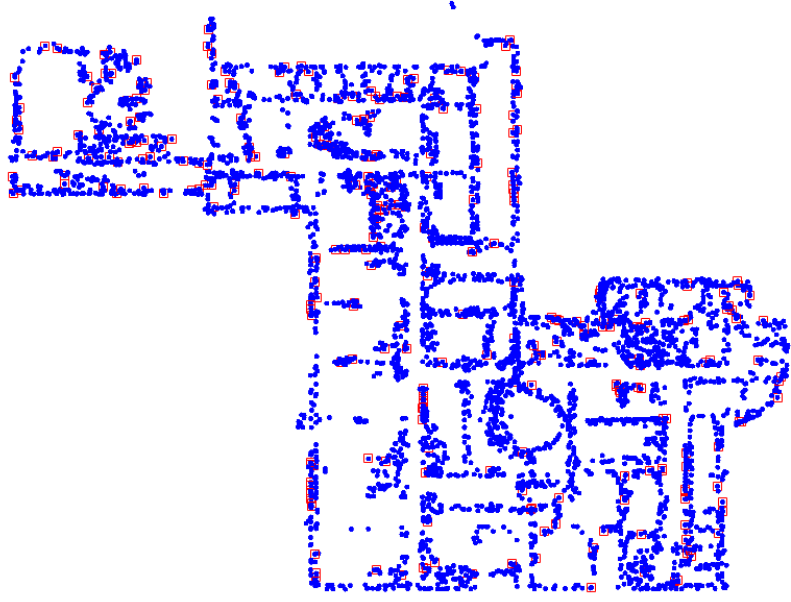}}}%
    \caption{Sub-figure (a) shows the layout of our environment consisting of an infinity number of reflector points which represent the set  $\mathcal{S}_{\epsilon}(\mathcal{M}_s)$. %For our simulations, we assume that for each test point, only a number of these reflector points are activated where their reflection coefficient has a uniform distribution between 0 and 1. 
    Sub-figure (b) shows the location of test points.}%%
    \label{map1}
\end{figure}
Fig. \ref{map1} (a) represents the physical layout of our environment extracted from the 4th floor of Bahen Centre at the University of Toronto. We assume that each of these black points is a reflector with a reflectivity coefficient of $\Gamma_i$, where $0 \leq \Gamma_i \leq 1$. Additionally, each time, based on the user's location, we activate a random number of reflector points on the boundary of the environment. We assume that at each user location, the BS can measure the AoA and ToA of different reception paths with variances $\sigma^2_{\theta}$ and $\sigma^2_{\tau}$, respectively. Furthermore, we assume that higher-order reflection paths have negligible effects on the received signal.

Our simulation has two major phases. In the first phase, we attempt to identify the environment by placing the user in specific locations and measuring AoA and ToA at the BS. Therefore, at each user location (test point), we can identify the location of some reflector points, typically having the same value as $n_r$, as defined in Section V. A higher value of $n_r$ leads to a quicker environmental identification. Based on the conclusion of the Lemma 4, for testing, we place the user's location close to the boundary of the physical layout to guarantee the identification of all effective reflector points in the environment. Fig. \ref{map1} (b) shows the locations of these test points, which are selected randomly with uniform distribution close to the physical boundary of the environment. As proved in Lemma 6, the following equation is an unbiased and asymptotically consistent estimator of $M_{\lambda}\left({\lambda_1,\lambda_2}\right)$ as $N$ approaches infinity:
\begin{equation}
\tilde{M}_{\lambda}(\lambda_1,\lambda_2)=\frac{\mathrm{Vol}(\mathcal{M}^{(\mathcal{R})}_s)}{N} \sum_{i=1}^{N}{e^{-2\pi j (\lambda_1 x_i+\lambda_2 y_i)}}.
\end{equation}
Since the value of $\mathrm{Vol}(\mathcal{M}^{(\mathcal{R})}_s)$ is not known, $\frac{1}{N} \sum_{i=1}^{N}{e^{-2\pi j (\lambda_1 x_i+\lambda_2 y_i)}}$ represents a scaled version of $\tilde{M}_{\lambda}(\lambda_1,\lambda_2)$.
\begin{figure}[!t]
  \centering
  \includegraphics[width=0.7\textwidth]{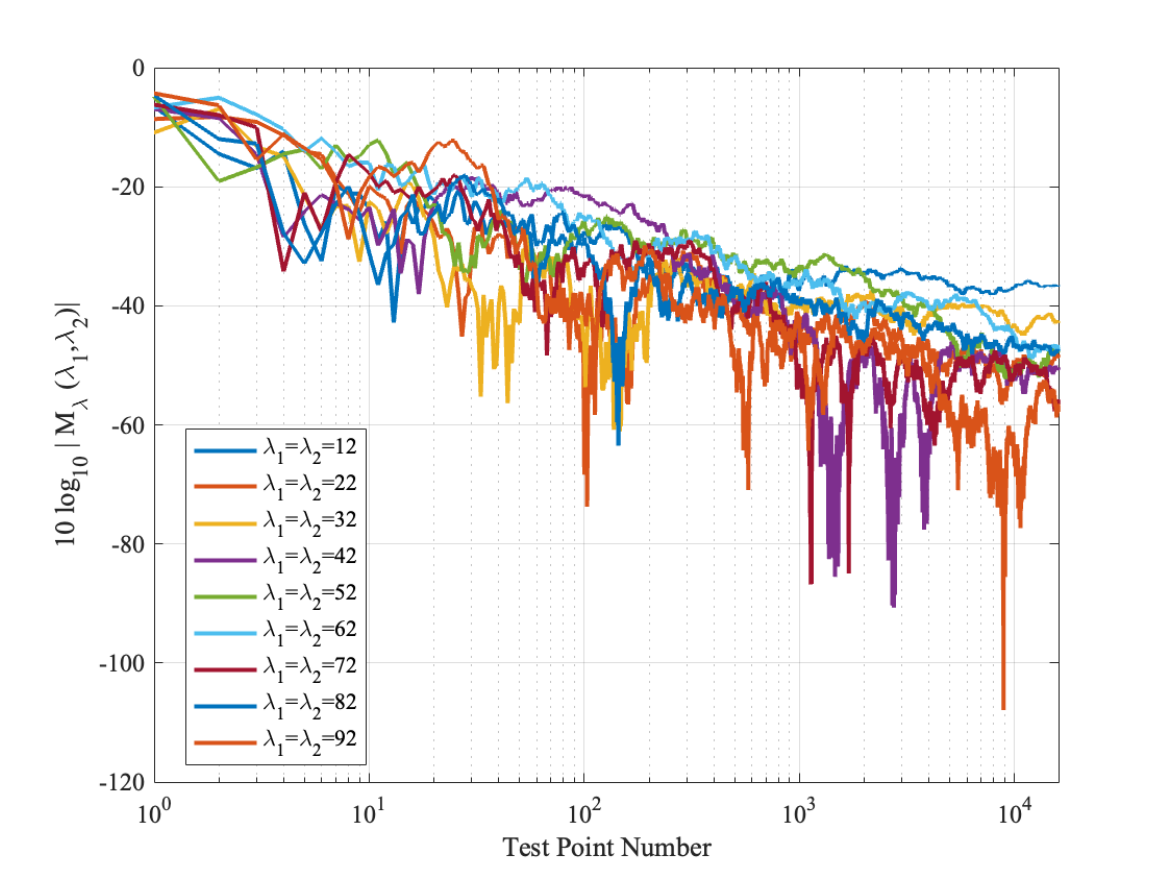}
  \caption{The value $10 \log_{10}\lvert M_{\lambda}(\lambda_1, \lambda_2)\rvert$ for different values of $\lambda_1=\lambda_2$ as a function of number of test points.} %As indicated, for a specific value of $N$, the value $10 \log_{10}\lvert M_{\lambda}(\lambda_1, \lambda_2)\rvert$ is sufficiently small or remains constant within a small range.}
  \label{map2a}
\end{figure}
For different values of $N$, Fig. \ref{map2a}  represents $10 \log_{10} {\lvert {\tilde{M}_{\lambda}(\lambda_1,\lambda_2)}\rvert}$ for specific cases of $\lambda_1=\lambda_2$ as a function of $N$. For $N>4000$ (the number of test points is roughly $N/n_r$), the value of $\lvert {\tilde{M}_{\lambda}(\lambda_1,\lambda_2)}\rvert$ exhibits minimal variations, indicating that additional test points are unnecessary, and we can construct the set of reflector objects with sufficient accuracy. Since in our simulation results, we consider $n_r=3$, the number of needed test points is close to $\frac{4 \times 10^3}{3}$. For our simulation, the area of the environment is $2\times 10^4$ square meters. For mm-wave applications where the wavelength is small, reaching $0.5 \mathrm{cm} \times 0.5 \mathrm{cm}$ precision requires at least $\frac{2 \times 10^4}{25 \times 10^{-4}}=8 \times 10^6$ test points (the actual number of needed test points is even higher, as in mm-wave applications, any change in location within half of the wavelength can drastically change the strength of the received signal). Comparing the minimum number of test points needed for the fingerprint-based approach, and the required number of test points for our approach, our technique in terms of the required number of test points is at least $6\times 10^4$ times more efficient than the fingerprint-based approach.
\color{black} 
\begin{figure}[!t]
  \centering
\includegraphics[width=0.7\textwidth]{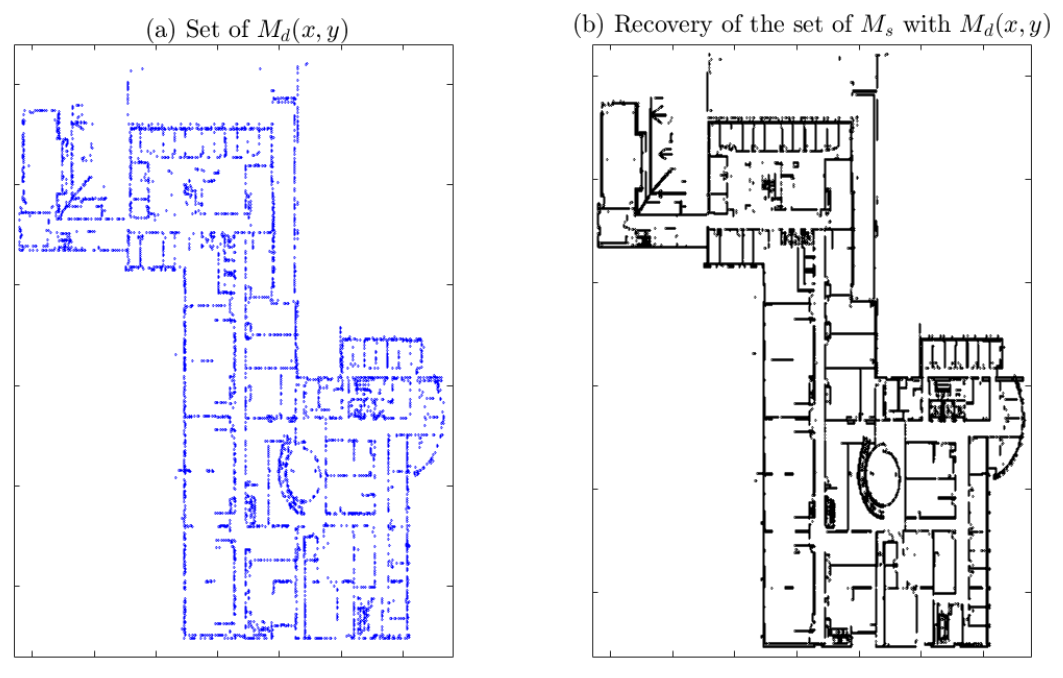}
  \caption{Sub-figure (a) represents the set of discrete samples  $\mathcal{M}_d(x,y)$ and sub-figure (b) shows the recovery version of the set $\mathcal{M}_s$ after using \eqref{recovery} with parameter $\alpha=0.2$ and iteration number of 10.}
  \label{recovery2a}
\end{figure}

Fig. \ref{recovery2a} shows the recovery of the set $\mathcal{M}_s$ from the sampling function $M_d(x,y)$ after 10 iterations and setting $\alpha=0.2$ in equation \eqref{recovery}. The results indicate that the recovered version of the set $\mathcal{M}_s$ closely aligns with the locations of reflector points in our environment with $\epsilon = 0.05$, coming from the definition of ``covering sheaf'' in Definition 5. 
\begin{figure}[!t]
  \centering
  \includegraphics[width=0.7\textwidth]{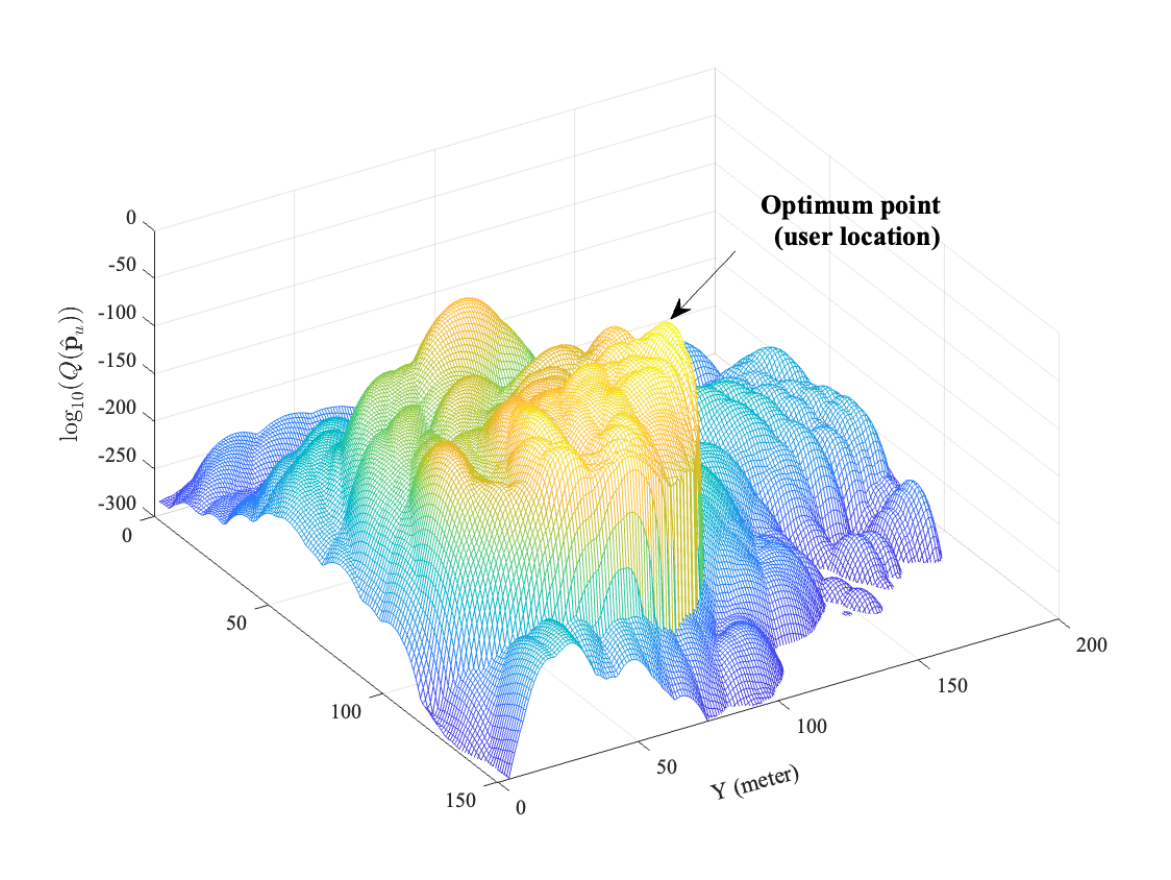}
  \caption{The value of the function $\log_{10}{Q\left({\hat{\bf p}_u}\right)}$, which is not convex. By pre-localization and limiting the search area we can speed up the optimization process. By focusing on a specific region rather than the entire domain, we can find optimal or near-optimal solutions more efficiently.}
  \label{cosh}
\end{figure}
Fig. \ref{cosh} shows the value of the function $\log_{10}{\left({Q(\hat{\bf p}_u)}\right)}$ versus $\hat{\bf p}_u$, indicating it is a non-convex function with multiple local maxima.
\begin{figure}[!t]
  \centering
\includegraphics[width=0.6\textwidth]{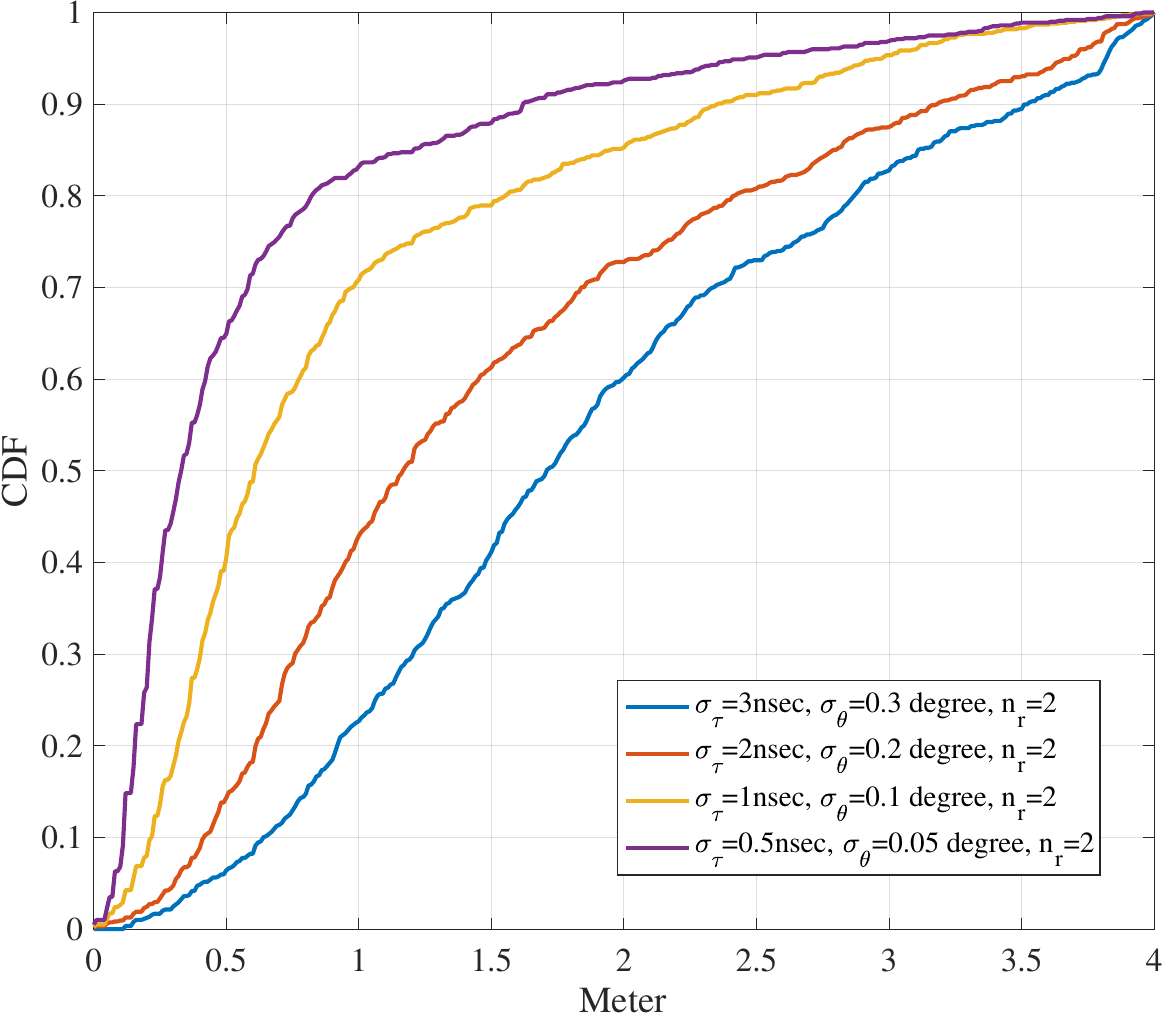}
  \caption{The Cumulative Distribution Function (CDF) measurement illustrates the performance of the proposed technique in an environment derived from the 4th floor of the Bahen Centre for Information Technology at the St. George campus of the University of Toronto.}
  \label{CDF}
\end{figure}
In Fig. \ref{CDF}, the Cumulative Distribution Function (CDF) measurement for our proposed technique has been conducted. We assume the receiver can measure AoA and ToA with different values of $\sigma_\theta$ and $\sigma_\tau$. The analysis encompasses various values of $\sigma_\tau$ and $\sigma_\theta$, representing the variances in time delay and angle measurements, respectively. In this scenario, the technique accounts for two typical first-order propagation paths\footnote{Since two reflection paths are insufficient to determine the user's location and a minimum of three paths is required, we assume that the receiver has access to a localization region spanning an area of $12m \times 12m$ to mitigate ambiguity in this scenario.}. The figure elucidates that as the variance of the measurements decreases, the accuracy of user location determination improves.
\begin{figure}[!t]
  \centering  \includegraphics[width=0.6\textwidth]{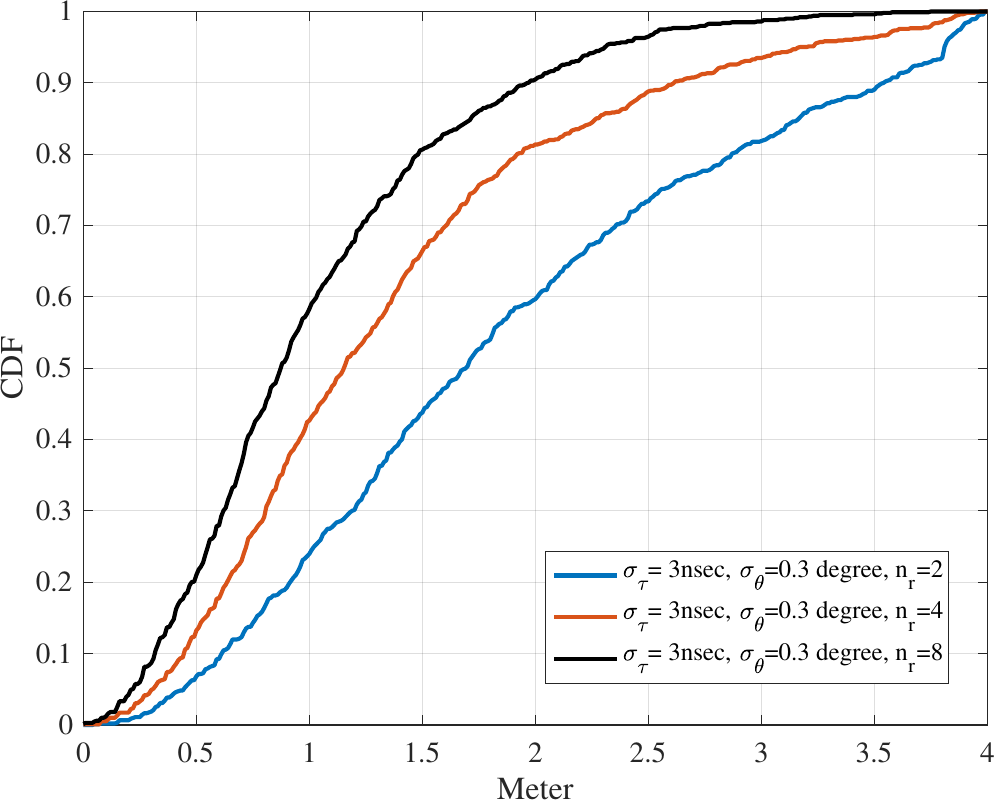}
  \caption{The CDF measurement demonstrates the performance of the proposed technique across varying numbers of typical first-order reflector points. It is evident that as the number of reflector points increases, the overall performance of the system improves.}
  \label{CDF2}
\end{figure}
Fig. \ref{CDF2} represents the value of the CDF for different numbers of reception paths where \(\sigma_\theta = 0.345^\circ\) and \(\sigma_\tau = 3 \, \text{nsec}\). The red curve depicts the CDF for the case where typically we engage 4 reflector points, while the black curve represents the CDF for the case where the number of reflection points is 8. This figure provides valuable insights, demonstrating that as the number of typical reflector points increases, consistent with the conclusion of the first theorem, our ability to accurately determine the user's location improves significantly.

In the following section of our simulation results, we explore various random physical structure environments. These environments feature randomly distributed reflector points across designated areas, ensuring that the coverage sheaf area (with $\epsilon=0.05$) remains constant relative to the total environmental area. Leveraging the high SNR at the BS, we assume that the BS can measure AoA and ToA with zero variance.
\begin{figure}[!t]
  \centering
\includegraphics[width=0.6\textwidth]{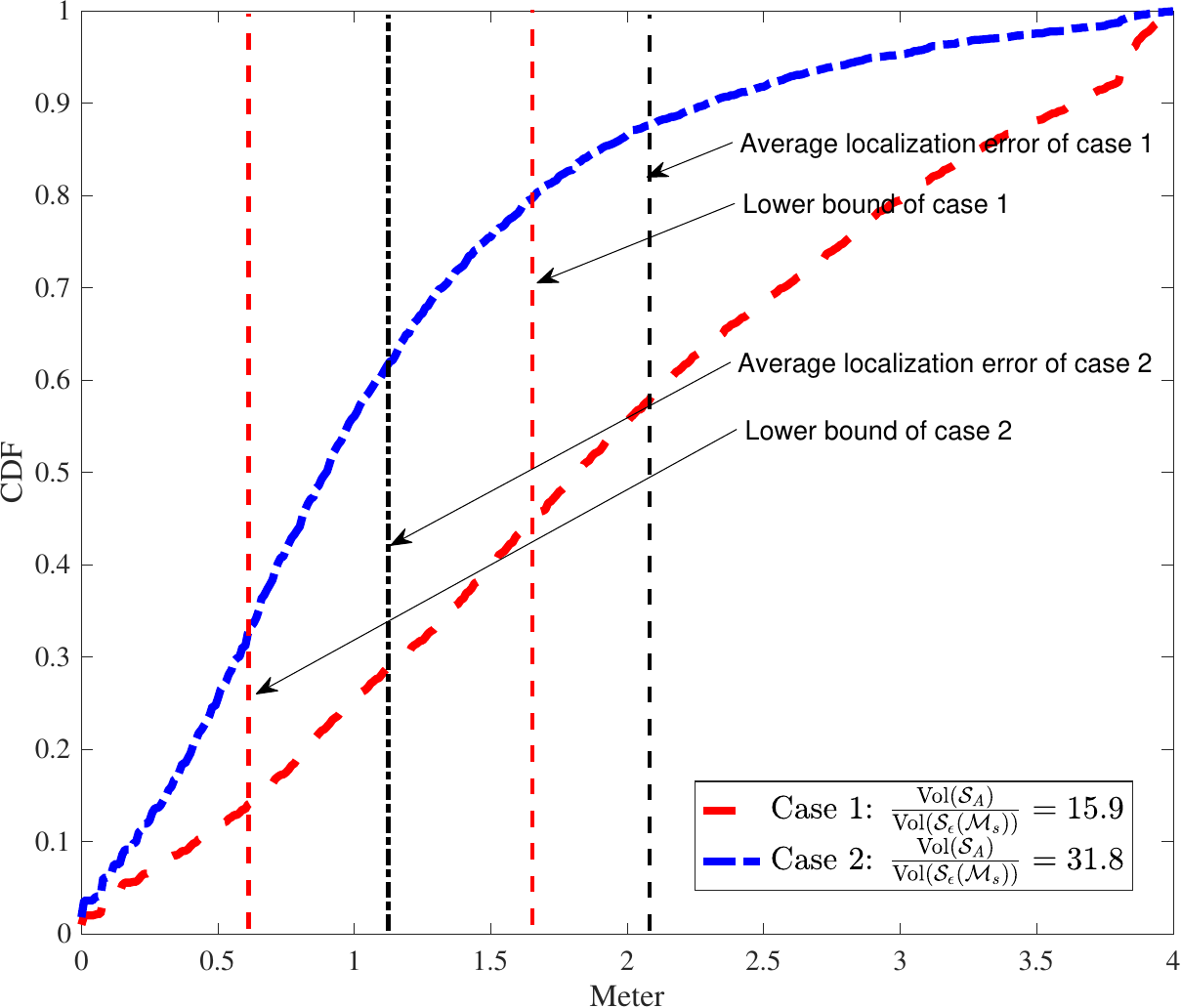}
  \caption{The CDF measurement demonstrates the performance of the proposed technique with $n_r=3$ for two different cases  $\frac{\mathrm{Vol}(\mathcal{S}_{A})}{\mathrm{Vol}(\mathcal{S}_\epsilon)}=15.9$ and $\frac{\mathrm{Vol}(\mathcal{S}_{A})}{\mathrm{Vol}(\mathcal{S}_\epsilon)}=31.8$ as $\mathrm{SNR}$ goes to infinity and $\epsilon=0.05$.}
  \label{CDF3l}
\end{figure}
Fig. \ref{CDF3l} depicts the CDF of the proposed technique for two distinct cases for both of these cases the SNR tends to infinity and $n_r=3$. In both cases, we examine 100 random physical structures for the reflector map, with the user randomly positioned more than $10^4$ times within a $200 \times 200$ square meter area. In the first case, the CDF reflects the localization technique when $\frac{\mathrm{Vol}\left({S_A}\right)}{\mathrm{Vol}\left(\mathcal{S}_{\epsilon}\left({\mathcal{M}_s}\right)\right)}=15.9$, while in the second scenario, this ratio is $31.8$. The lower-bound of $\mathrm{Vol}\left({\mathcal{S}_u}\right)$ is $2.6$ square meters in the first scenario and $0.36$ square meters in the second. Correspondingly, for a circular ambiguity area, these values translate to radii of $1.6$ and $0.6$ meters, respectively.\color{black}
 \section{Conclusion}
\label{sec:conclusion}
 We introduced a localization strategy tailored for indoor environments, harnessing the power of reflection points to ascertain the user's position. Our strategy unfolds in two pivotal stages, commencing with an expedited environmental sensing phase designed to construct a map of fixed reflector locations, notably reducing the number of test points compared to conventional fingerprint-based techniques. One of the results of this work lies in the development of a theorem that establishes an upper bound on the Log-scale accuracy ratio, particularly in high signal-to-noise ratio scenarios. This theoretical framework not only enhances our understanding of the strategy's performance limits but also provides a benchmark for assessing its efficacy across diverse environmental parameters. Through the formulation of a maximization problem that considers the likelihood of each set of fixed reflectors, we achieve an estimation of the user's location. Beyond theoretical advancements, our simulation results provide tangible evidence of the strategy's efficacy, underlining its versatility.
%\section*{Acknowledgement}
%...
\appendices
\section{Proof of first lemma and computing the covariance matrix  ${\bf V}$}
  \label{appendix1}
In polar coordinates, let the BS at one focus, and the user has the angular coordinate  $\phi$ to BS and the x-axis. In this case, the ellipse's equation is:
\begin{equation}
\label{elippolar}
r(\theta)=\frac{a(1-e^2)}{1-e \cos (\theta-\phi)},~e=\frac{c}{a},
\end{equation}   
where $b$ and $a={c_0 \tau}/{2}$ are the width and length of the ellipse, respectively and $c=\sqrt{a^2-b^2}={d_{ub}}/{2}$ where $d_{ub}=\vert{{\bf p}_u-{\bf p}_b}\vert$. In addition, $0 \leq \theta \leq 2\pi$ represents the polar angle measured with respect to the x-axis. Therefore, \eqref{elippolar} can be represented as follows:
\begin{equation}
r(\theta,\tau) = \frac{c^2_0 \tau^2 - d^2_{ub}}{2 \left(c_0 \tau - d_{ub} \cos (\theta-\phi)\right)}.
\end{equation}
In the above equation, the function $r(\theta)$ has been modified to $r(\theta,\tau)$ because the ellipse parameters are directly dependent on $a$, which is, in turn, influenced by the propagation time $\tau$. The physical location of the reflector as a function  $\theta$ and $\tau$ can be calculated as follows:
\begin{equation}
x_s=r(\theta,\tau) \cos(\theta),~y_s=r(\theta,\tau) \sin(\theta).
\end{equation}
A small variation in the measured values  $\tau$ and $\theta$ within the values  $\pm \Delta \tau$ and $\pm \Delta \theta$ respectively leads to the following variations on the measured position of the reflector on the x and y-axis as follows:
\begin{align}
\Delta x_s&\approx \frac{\partial \left[ r(\theta,\tau) \cos(\theta)\right]}{\partial \theta} \Delta \theta+\frac{\partial \left[ r(\theta,\tau) \cos(\theta) \right]}{\partial \tau} \Delta \tau\\
\Delta y_s&\approx \frac{ \partial \left[ r(\theta,\tau) \sin(\theta) \right]}{\partial \theta} \Delta \theta+\frac{\partial \left[ r(\theta,\tau) \sin(\theta) \right]}{\partial \tau} \Delta \tau.
\end{align}
In order to compress our notation we replace $r(\theta,\tau)$ with $r$. Therefore, we have:
\begin{equation}
\label{dx}
\Delta x_s \approx \left({\frac{\partial r}{\partial \theta} \cos(\theta)- r \sin(\theta)}\right) \Delta \theta + \frac{\partial r}{\partial \tau} \cos(\theta) \Delta \tau.
\end{equation} 
Similarly, the measured position of the reflector on the y-axis can be calculated as follows:
\begin{equation}
\label{dy}
\Delta y_s \approx \left({\frac{\partial r}{\partial \theta} \sin(\theta)+ r \cos(\theta)}\right) \Delta \theta + \frac{\partial r}{\partial \tau} \sin(\theta) \Delta \tau,
\end{equation} 
where indicates that $\mathbb{E}\{\Delta x_s\}=0$ and $\mathbb{E}\{\Delta y_s\}=0$. Considering the variances  $\sigma^2_{\theta}$ and $\sigma^2_{r}$ for both variables $\theta$ and $r$, respectively where both of them have small enough value. The value  $\mathbb{E}\{{\Delta x ^2}\}=\sigma^2_x$ and $\mathbb{E}\{{\Delta y ^2}\}=\sigma^2_y$ can be calculated as follows:
\small
\begin{align}
\sigma^2_{x_s} \approx &\left({\frac{\partial r}{\partial \theta} \cos(\theta)- r \sin(\theta)}\right)^2 \sigma^2_\theta + \left({\frac{\partial r}{\partial \tau} \cos(\theta)}\right)^2 \sigma^2_\tau.\\
\sigma^2_{y_s} \approx &\left({\frac{\partial r}{\partial \theta} \sin(\theta)+ r \cos(\theta)}\right)^2 \sigma^2_{\theta} + \left(\frac{\partial r}{\partial \tau} \sin(\theta)\right)^2 \sigma^2_\tau.
\end{align}
\normalsize
Similarly, $\mathbb{E}\{{\Delta x \Delta y}\}=\sigma_{x_s y_s}$ can be calculated by finding expectation value on the multiplication of equations \eqref{dx} and \eqref{dy} as follows:
\small
\begin{align}
\sigma_{x_s y_s}&=\left({\frac{1}{2}{\left({\left(\frac{\partial r}{\partial \theta}\right)^2}-r^2\right)}+\frac{\partial r}{\partial \theta} r \cos{(2\theta)}}\right)\sigma^2_{\theta}\\
&+\frac{1}{2}\left({\frac{\partial r}{\partial \tau}}\right)^2 \sin{(2 \theta)} \sigma^2_{\tau}\\
&+\left({\frac{1}{2} \frac{\partial^2 r}{\partial \theta \partial \tau} \sin(2 \theta)+\frac{\partial r}{\partial \tau} r \cos^2 (\theta)-\frac{\partial r}{\partial \theta} r \sin^2(\theta)}\right) \sigma_{\theta,\tau},
\end{align}
\normalsize
where since $\theta$ and $\tau$ are two independent random variables the last term  $\sigma_{\theta,\tau}$ has zero value, finally we have:
\begin{equation}
{\bf V}_s=\begin{bmatrix}
\sigma^2_{x_s} & \sigma_{x_s y_s} \\
\sigma_{y_s x_s} & \sigma^2_{y_s}
\end{bmatrix}.
\end{equation}

 \section{Extracting $\mathcal{M}^{(\mathcal{R})}_s$ from the discrete set $\mathcal{M}^{(\mathcal{R})}_d$}
 \label{recoverysetMsfromMd}
 The set of random samples extracted from the set  $\mathcal{M}^{(\mathcal{R})}_s$ can be represented as
\begin{equation}
M_d(x,y)=\sum_{i}{\delta(x-x_i,y-y_i)},
\end{equation}
where in the above relation $x_i \in \mathcal{X}$ and $y_i \in \mathcal{Y}$, represent the sampling points of the function $M(x,y)$ and $(x_i,y_i) \in \mathcal{M}_d \subseteq \mathcal{M}^{(\mathcal{R})}_s$. % and for simplicity of notation this operation is represented by $g_d(x,y)=S g(x,y)$.
Let us define the 2-dimensional function $l(x,y)$ as follows:
\begin{equation}
l(x,y)={\sin{(2 \pi \lambda_{\mathrm{m}}x)}\sin{(2 \pi \lambda_{\mathrm{m}}y)}}/{\pi^2 xy},
\end{equation}
with the following 2-dimensional Fourier transform:
\begin{equation}
L(\lambda_1,\lambda_2)=\left\{ 
  \begin{array}{ c l }
    1 & \quad \textrm{if } \lvert \lambda_i \rvert < \lambda_{\mathrm{m}},~i=1,2 \\
    0                 & \quad \textrm{if } \lvert \lambda_i \rvert > \lambda_{\mathrm{m}},~i=1,2
  \end{array}.
\right.
\end{equation}
It has been shown in \cite{Marvasti1} that $M(x,y)$ can be extracted by following an iterative method from $M^{(k)}(x,y)$ as
\begin{align}
\label{recovery}
M^{(k+1)}&(x,y)= \alpha l(x,y)*{M}_d(x,y)+l(x,y)*M^{(k)}(x,y)\nonumber\\ 
&-\alpha l(x,y)*\sum_{i}{M^{(k)}(x,y)\delta(x-x_i,y-y_i)},
\end{align}
where $\alpha$ is convergence constant and the function  $l(x,y)*M(x,y)$ can be simplified as follows:
\small
\begin{align}
l(x,y)*M_d(x,y)&={\iint_{-\infty}^{\infty}{l(u,v)M(x-u,y-v)du dv}}\\
&=\sum_{i}{l(x-x_i,y-y_i)}.
\end{align}
\normalsize
 For specific range of parameter $\alpha$, it has been shown in \cite{Marvasti1, Ignjatović, Oppenheim}, if $\| M^{(k+1)}(x,y) - M^{(k)}(x,y) \| \leq \| M^{(k)}(x,y) - M^{(k-1)}(x,y) \|$ then we have:
\begin{equation}
\lim_{k \rightarrow \infty}{M^{(k)}(x,y)} = M(x,y).
\end{equation}
 %Figure.~\ref{Fig8} represents the value  $D^{(t)}({{\bf p}_u})$ for different iterations, where the reflection coefficient of the third user is 0.01. 
 Finally, the BS estimates the location of reflector and scatterer objects, which we refer to as the ``reflection map'' of the environment. This algorithm empowers us to construct a map that illustrates the area containing the locations of scatterer and reflector objects within the environment with probability at least $1-\epsilon$. In the rest of the paper, for simplicity in notation, we omit the index  $(\mathcal{R})$ from $\mathcal{M}^{(\mathcal{R})}_s$.
 \section{Proof of Lemma 5}
 \label{proofunbiased}
 Starting from finding the expectation value of \eqref{disf}, we have:
\footnotesize
\begin{align}
\mathbb{E}&\Big\lbrace{\frac{\mathrm{Vol}\left({\mathcal{M}^{(\mathcal{R})}_s}\right)}{N}\sum_{i=1}^{N}{e^{-2\pi j(\lambda_1 x_i+\lambda_2 y_i)}}}\Big\rbrace \nonumber\\
&=\frac{\mathrm{Vol}\left({\mathcal{M}^{(\mathcal{R})}_s}\right)}{N}\sum_{i=1}^{N}{\iint_{\mathcal{M}^{(\mathcal{R})}_s}{\frac{1}{\mathrm{Vol}\left({\mathcal{M}^{(\mathcal{R})}_s}\right)}e^{-2\pi j(\lambda_1 x_i+\lambda_2 y_i)} dx_i dy_i}}\\
&=\frac{\mathrm{Vol}\left({\mathcal{M}^{(\mathcal{R})}_s}\right)}{N}\sum_{i=1}^{N}{\frac{1}{\mathrm{Vol}\left({\mathcal{M}^{(\mathcal{R})}_s}\right)} M_{\lambda}(\lambda_1,\lambda_2)}\\
&=M_{\lambda}(\lambda_1,\lambda_2).
\end{align}
\normalsize
Above relations show that $\tilde{M}_{\lambda}(\lambda_1,\lambda_2)$ is an unbiased estimator  $M_{\lambda}(\lambda_1,\lambda_2)$. The variance of the estimator $\tilde{M}_{\lambda}(\lambda_1,\lambda_2)$ can be calculated as follows:
\begin{align}
\mathbb{E}&\left\{{\lvert \tilde{M}_{\lambda}(\lambda_1,\lambda_2) \rvert}^2\right\} - \lvert \mathbb{E}\{\tilde{M}_{\lambda}(\lambda_1,\lambda_2)\} \rvert^2 \nonumber\\
&= \mathbb{E}\left\{{\lvert \tilde{M}_{\lambda}(\lambda_1,\lambda_2) \rvert}^2\right\} - \lvert M_{\lambda}(\lambda_1,\lambda_2) \rvert^2.
\end{align}
\begin{figure*}%[u]
%\hrule
\small
\begin{align}
\label{varspec1}
\mathbb{E}\{{\lvert{\tilde{M}_{\lambda}(\lambda_1,\lambda_2)}\rvert}^2\}&=\left({\frac{\mathrm{Vol}(\mathcal{M}^{(\mathcal{R})}_s)}{N}}\right)^2 \mathbb{E}\Big\{{\sum_{i=1}^{N}{e^{-2 \pi j (\lambda_1 x_i + \lambda_2 y_i)}}}{\sum_{k=1}^{N}{e^{2 \pi j (\lambda_1 x_k + \lambda_2 y_k)}}}\Big\}\\
%&=\left({\frac{\mathrm{Vol}(\mathcal{M}^{(\mathcal{R})}_s)}{N}}\right)^2 \mathbb{E}\Big\{{\sum_{i=1}^{N}{1}}+{\sum_{\substack{k,i=1\\ k \neq i}}^{N}{e^{-2 \pi j (\lambda_1 (x_i-x_k) + \lambda_2 (y_i-y_k))}}}\Big\}\\
\label{varspec2}
&\overset{(a)}{=}\left({\frac{\mathrm{Vol}(\mathcal{M}^{(\mathcal{R})}_s)}{N}}\right)^2 \Bigg(N+{\sum_{\substack{k,i=1\\ k \neq i}}^{N}{ \mathbb{E}\Big\{e^{-2 \pi j (\lambda_1 x_i + \lambda_2 y_i)}\Big\}\mathbb{E}\Big\{e^{2 \pi j (\lambda_1 x_k + \lambda_2 y_k)}\Big\}}}\Bigg)\\
\label{varspec3}
&=\left({\frac{\mathrm{Vol}(\mathcal{M}^{(\mathcal{R})}_s)}{N}}\right)^2 \Bigg({N+{\sum_{\substack{k,i=1\\ k \neq i}}^{N}{{\frac{\lvert M_{\lambda}(\lambda_1,\lambda_2) \rvert^2}{\mathrm{Vol}(\mathcal{M}^{(\mathcal{R})}_s)^2}}}}}\Bigg)=\frac{\mathrm{Vol}(\mathcal{M}^{(\mathcal{R})}_s)^2}{N}+\frac{N(N-1) \lvert M_{\lambda}(\lambda_1,\lambda_2) \rvert^2}{N^2}
\end{align}
\hrule
\normalsize
\end{figure*}
In the above relation the term  $\mathbb{E}\{{\lvert{\tilde{M}_{\lambda}(\lambda_1,\lambda_2)}\rvert}^2\}$ can be simplified as in relation \eqref{varspec1} to \eqref{varspec3}.
In equation \eqref{varspec3}, (a) stems from the fact that $f_R({\bf p})=\frac{1}{\mathrm{Vol}\left({\mathcal{M}^{(\mathcal{R})}_s}\right)}$, and every pair of tuples $(x_i,y_i)$ and $(x_k,y_k)$ are independent. Finally, the variance of the estimator can be calculated as follows:
\begin{align}
\mathbb{E}{{\lvert{\tilde{M}_{\lambda}(\lambda_1,\lambda_2)}\rvert}^2}&-\lvert \mathbb{E} \{\tilde{M}_{\lambda}(\lambda_1,\lambda_2)\}  \rvert^2\nonumber\\
&=\frac{\mathrm{Vol}(\mathcal{M}^{(\mathcal{R})}_s)-\lvert {M}_{\lambda}(\lambda_1,\lambda_2) \rvert^2}{N}.
\end{align}
The above relation indicates that as the number of random samples increases, the variance of the estimated spectrum approaches zero, converging to the value  $M_{\lambda}(\lambda_1, \lambda_2)$.  
 \section{Fourier Representation  $f_S ({\bf p}) M(x,y)$}
 \label{FourierRepresentation}
 We use following estimator to find ${M}_{\lambda}(\lambda_1,\lambda_2)=\mathcal{F}\left({M(x,y) f_S ({\bf p})}\right)$:
\begin{equation}
\label{disf2}
\tilde{M}_{\lambda}(\lambda_1,\lambda_2)=\frac{1}{N}\sum_{i=1}^{N}{ e^{-2\pi j(\lambda_1 x_i+\lambda_2 y_i)}}.
\end{equation}
Similar to the uniform sampling, we show that the above estimator has zero bias and an asymptotically consistent estimator, and it converges to the true value since by increasing $N$ the variance goes to zero value. Therefore, similar to the previous approach, we consider $N$ reflector points distributed with the distribution  $f_S ({\bf p})$ across the area set  $\mathcal{S}_{\epsilon}(\mathcal{M}_s)$, the expectation value of \eqref{disf2} can be calculated as follows:
\small
\begin{align}
\mathbb{E}&\Big\lbrace{\frac{1}{N}\sum_{i=1}^{N}{e^{-2\pi j(\lambda_1 x_i+\lambda_2 y_i)}}}\Big\rbrace \nonumber\\
&=\frac{1}{N}\sum_{i=1}^{N}{\iint_{\mathcal{S}_{\epsilon}(\mathcal{M}_s)}{f_S({\bf p}) e^{-2\pi j(\lambda_1 x_i+\lambda_2 y_i)} dx_i dy_i}}\\
&=\frac{1}{N}\sum_{i=1}^{N}{\iint_{S}{f_S({\bf p})M(x,y) e^{-2\pi j(\lambda_1 x_i+\lambda_2 y_i)} dx_i dy_i}}\\
&=\frac{1}{N}\sum_{i=1}^{N}{\mathcal{F}\left({f_S({\bf p}) M(x,y)}\right)}\\
&=\mathcal{F}\left({f_S({\bf p}) M(x,y)}\right).
\end{align}
\normalsize
Above relations shows that $\tilde{M}_{\lambda}(\lambda_1,\lambda_2)$ is unbiased estimator  $M_{\lambda}(\lambda_1,\lambda_2)$. The variance of the estimator $\tilde{M}_{\lambda}(\lambda_1,\lambda_2)$ can be calculated as follows:
\begin{align}
\mathbb{E}&\left\{ \lvert \tilde{M}_{\lambda}(\lambda_1, \lambda_2) \rvert^2 \right\} - \lvert \mathbb{E}\{ \tilde{M}_{\lambda}(\lambda_1, \lambda_2) \} \rvert^2 \nonumber \\
&= \mathbb{E}\left\{ \lvert \tilde{M}_{\lambda}(\lambda_1, \lambda_2) \rvert^2 \right\} - \lvert M_{\lambda}(\lambda_1, \lambda_2) \rvert^2.
\end{align}
The term  $\mathbb{E}\{{\lvert{\tilde{M}_{\lambda}(\lambda_1,\lambda_2)}\rvert}^2\}$ can be simplified as follows:
\small
\begin{align}
&\mathbb{E}\{{\lvert{\tilde{M}_{\lambda}(\lambda_1,\lambda_2)}\rvert}^2\}\nonumber\\
&=\frac{1}{N^2} \mathbb{E}\Big\{{\sum_{i=1}^{N}{e^{-2 \pi j (\lambda_1 x_i + \lambda_2 y_i)}}}{\sum_{k=1}^{N}{e^{2 \pi j (\lambda_1 x_k + \lambda_2 y_k)}}}\Big\}\\
%&=\frac{1}{N^2} \mathbb{E}\Big\{{\sum_{i=1}^{N}{1}}+{\sum_{\substack{k,i=1\\ k \neq i}}^{N}{e^{-2 \pi j (\lambda_1 (x_i-x_k) + \lambda_2 (y_i-y_k))}}}\Big\}\\
&\overset{(a)}{=}\frac{1}{N^2} \Bigg(N+{\sum_{\substack{k,i=1\\ k \neq i}}^{N}{ \mathbb{E}\Big\{e^{-2 \pi j (\lambda_1 x_i + \lambda_2 y_i)}\Big\}\mathbb{E}\Big\{e^{2 \pi j (\lambda_1 x_k + \lambda_2 y_k)}\Big\}}}\Bigg)\\
&=\frac{1}{N^2} \Bigg({N+{\sum_{\substack{k,i=1\\ k \neq i}}^{N}{\lvert {M_{\lambda}(\lambda_1,\lambda_2)}\rvert ^2}}}\Bigg)\\
&=\frac{1+(N-1) \lvert M_{\lambda}(\lambda_1,\lambda_2) \rvert^2}{N}
\end{align}
\normalsize
where (a) comes from the fact that $f_R({\bf p})=\frac{1}{\mathrm{Vol}(\mathcal{S}_{\epsilon}(\mathcal{M}_s))}$ and every two tuples  $(x_i,y_i)$ and $(x_k,y_k)$ are independents. The variance of the estimator can be calculated as follows:
\begin{align}
\mathbb{E}\{{\lvert{\tilde{M}_{\lambda}(\lambda_1,\lambda_2)}\rvert}^2\}&-\lvert \mathbb{E}\{\tilde{M}_{\lambda}(\lambda_1,\lambda_2)\} \rvert^2\nonumber\\
&=\frac{\mathrm{Vol}(\mathcal{S}_{\epsilon}(\mathcal{M}_s))-\lvert {M}_{\lambda}(\lambda_1,\lambda_2) \rvert^2}{N}.
\end{align}
The above relation indicates that as the number of random samples increases, the variance of the estimated spectrum approaches zero, converging to the value  $M_{\lambda}(\lambda_1,\lambda_2)$.
\section{Proof of Theorem 1}
\label{appendix0}
\color{black} Let ${\bf{p}}_u$ denote the location of a transmitter. We introduce the variable $K = \frac{\mathrm{Vol}(\mathcal{S}_{A})}{\mathrm{Vol}(\mathcal{S}_u)}$, representing the number of possible candidate cells for the user's location in the set of partitions $\mathcal{PL} = \{ \mathcal{P}_1,\dots,\mathcal{P}_K \}$, ${\mathcal{P}_i} \cap {\mathcal{P}_j} = \varnothing$ and $\cup_{i}{\mathcal{P}_i} = \mathcal{S}_A$, where $\mathcal{S}_{A}$ represents the accessible area where a user can move without physical limitation. 
Only one of these partitions (cells) accurately corresponds to the user's position, matching the location of the user ${\bf{p}}_u$, we assume that ${\bf{p}}_u \in \mathcal{P}_1$.
\begin{definition}
A {\it``Mapped partition''} $\mathcal{P}'\left(\tau_i,\theta_i, \mathcal{P}_j\right)$ is a set of points defined as follows:
\begin{align}
\mathcal{P}'&\left(\tau_i,\theta_i, \mathcal{P}_j\right)\nonumber\\
&=\Big\{(x,y) \in \mathbb{R}^2 : (x,y)=\mathcal{J}\left(\tau_i, \theta_i, {\bf p}\right), \forall {\bf p} \in \mathcal{P}_j\Big\}.
\end{align}
\end{definition}
The error occurs if we find a location ${\bf{p}'} \notin \mathcal{P}_1$ that maps the entire measurement set $\mathcal{M}_e$ using the function $\mathcal{J}\left({(\tau_i,\theta_i), {\bf p}'}\right)$ into the covering sheaf $\mathcal{S}_{\epsilon}(\mathcal{M}_s)$. We can define the $\mathcal{E}_{ji}$ error event as follows:
\begin{equation}
\mathcal{E}_{ji}=\Big\{\exists {\bf{p}}' \in \mathcal{P}_j \neq \mathcal{P}_1: \mathcal{J}\left({(\tau_i,\theta_i), {\bf p}' }\right) \in \mathcal{S}_{\epsilon}({\mathcal{M}_s})\Big\}.
\end{equation}
In a scenario where the $\mathrm{SNR}$ tends towards infinity, the receiver can extract $\tau_i$ and $\theta_i$ with zero variances. Consequently, the mapped partition $\mathcal{P}'\left(\tau_i,\theta_i, \mathcal{P}_j\right)$ forms a segment with an angle of $\theta_i$ corresponding to the receiver. Assuming each partition has a small size, we can upper-bound the maximum length of this segment as follows:
\begin{equation}
D\left({\mathcal{P}'\left(\tau_i,\theta_i, \mathcal{P}_j\right)}\right) \leq D\left({\mathcal{P}_j}\right) \max_{{{\bf p} \in \mathcal{P}_j}} \Bigg \lvert {\frac{\partial \mathcal{J}\left({(\tau_i,\theta_i), {\bf p} }\right) }{ \partial {\bf p}}} \Bigg \rvert,
\end{equation}
where for every partition $\mathcal{P}$, we have:
\begin{equation}
D(\mathcal{P})=\max_{{\bf p}_1, {\bf p}_2 \in \mathcal{P}}{\lvert {\bf p}_1 - {\bf p}_2 \rvert}.
\end{equation}
If ${\bf p} = x_p \hat{a}_x + y_p \hat{a}_y$, we have:
\small
\begin{equation}
\Big \lvert {\frac{\partial \mathcal{J}\left({(\tau_i,\theta_i), {\bf p} }\right) }{ \partial {\bf p}}} \Big \rvert = \Big \lvert {\frac{\partial \mathcal{J}\left({(\tau_i,\theta_i), {\bf p} }\right) }{ \partial {x_p}}\hat{a}_x+\frac{\partial \mathcal{J}\left({(\tau_i,\theta_i), {\bf p} }\right) }{ \partial {y_p}}} \hat{a}_y \Big \rvert.
\end{equation}
\normalsize
The probability that the mapped partition of $\mathcal{P}_j$ does not intersect with the set $\mathcal{S}_{\epsilon}(\mathcal{M}_s)$ can be lower-bounded as follows:
\small
\begin{align}
\label{1pro}
\mathbb{P}&({\mathcal{P}'\left(\tau_i,\theta_i, \mathcal{P}_j\right) \notin \mathcal{S}_{\epsilon}(\mathcal{M}_s)}) \geq \nonumber\\ &\frac{\mathrm{Vol}(\mathcal{S}_{A})-\max_{i,j} D\left({\mathcal{P}'\left(\tau_i,\theta_i, \mathcal{P}_j\right)}\right) l_{\mathcal{S}_{\epsilon}(\mathcal{M}_s)}}{\mathrm{Vol}(\mathcal{S}_{\epsilon}(\mathcal{M}_s)) + \mathrm{Vol}(\mathcal{S}_{A})},
\end{align}
\normalsize
where $l_{\mathcal{S}_{\epsilon}(\mathcal{M}_s)}$ represents the length of the side of the covering sheaf. The term $\max_{i,j} D\left({\mathcal{P}'\left(\tau_i,\theta_i, \mathcal{P}_j\right)}\right) l_{\mathcal{S}_{\epsilon}(\mathcal{M}_s)}$ arises from the fact that if we have an empty strip with the same length as the side of the covering sheaf and with a thickness of at least $\max_{i,j} D\left({\mathcal{P}'\left(\tau_i,\theta_i, \mathcal{P}_j\right)}\right)$, it guarantees that if the starting point of the segment line $\mathcal{P}'\left(\tau_i,\theta_i, \mathcal{P}_j\right)$ is outside the strip, it will never intersect the covering sheaf. The numerator in the above relation represents the area of the accessible area minus the guard strip with a thickness of $\max_{i,j} D\left({\mathcal{P}'\left(\tau_i,\theta_i, \mathcal{P}_j\right)}\right)$, and the denominator represents the total area of the localization procedure. Therefore, from \eqref{1pro}, we can conclude that:
\small
\begin{align}
\mathbb{P}&({\mathcal{P}'\left(\tau_i,\theta_i, \mathcal{P}_j\right) \in \mathcal{S}_{\epsilon}(\mathcal{M}_s)})\nonumber\\
&\leq \frac{\mathrm{Vol}(\mathcal{S}_{\epsilon}(\mathcal{M}_s))+\max_{i,j} D\left({\mathcal{P}'\left(\tau_i,\theta_i, \mathcal{P}_j\right)}\right) l_{\mathcal{S}_{\epsilon}(\mathcal{M}_s)}}{\mathrm{Vol}(\mathcal{S}_{\epsilon}(\mathcal{M}_s)) + \mathrm{Vol}(\mathcal{S}_{A})}\\
& \approx \frac{\mathrm{Vol}(\mathcal{S}_{\epsilon}(\mathcal{M}_s))}{\mathrm{Vol}(\mathcal{S}_{\epsilon}(\mathcal{M}_s)) + \mathrm{Vol}(\mathcal{S}_{A})}.
\end{align}
\normalsize
The last stage arises from the fact that, for small sizes of partitions, $\max_{i,j} D\left({\mathcal{P}'\left(\tau_i,\theta_i, \mathcal{P}_j\right)}\right)$ approaches zero. Finally, for the $j$-th partition, an error will occur if we can find ${\bf p} \in \mathcal{P}_j \neq \mathcal{P}_1$, where for all $\left({\tau_i,\theta_i}\right) \in \mathcal{M}_e$, the function $\mathcal{J}(\tau_i,\theta_i,{\bf p})$ falls within the set of covering sheaf, represented by $\mathbb{P}\left({\mathcal{E}_j}\right)$. This can be calculated as follows:
\begin{align}
\mathbb{P}\left({\mathcal{E}_j}\right)&=\mathbb{P}\left({{\mathcal{P}'\left(\tau_i,\theta_i, \mathcal{P}_j\right) \in \mathcal{S}_{\epsilon}(\mathcal{M}_s)}, 1 \leq i \leq n_r}\right)\\
&= \prod_{i=1}^{n_r}{\frac{\mathrm{Vol}(\mathcal{S}_{\epsilon}(\mathcal{M}_s))}{\mathrm{Vol}(\mathcal{S}_{\epsilon}(\mathcal{M}_s)) + \mathrm{Vol}(\mathcal{S}_{A})}}.
\end{align}
Therefore, total error probability $\mathbb{P}\left({\mathcal{E}=\bigcup_{i,j} \mathcal{E}_{ij}}\right)$ can be calculated as follows:
\small
\begin{align}
\mathbb{P}\left({\mathcal{E}}\right)&=1-\left({1-\left(\frac{\mathrm{Vol}(\mathcal{S}_{\epsilon}(\mathcal{M}_s))}{\mathrm{Vol}(\mathcal{S}_{\epsilon}(\mathcal{M}_s)) + \mathrm{Vol}(\mathcal{S}_{A})}\right)^{n_r}}\right)^{\left({K-1}\right)}\\
&\approx (K-1)\left(\frac{\mathrm{Vol}(\mathcal{S}_{\epsilon}(\mathcal{M}_s))}{\mathrm{Vol}(\mathcal{S}_{\epsilon}(\mathcal{M}_s)) + \mathrm{Vol}(\mathcal{S}_{A})}\right)^{n_r}.
\end{align}
\normalsize
For a sufficiently large value $K = \frac{\mathrm{Vol}(\mathcal{S}_{A})}{\mathrm{Vol}(\mathcal{S}_u)}$, the value $\mathbb{P}\left({\mathcal{E}}\right)$ tends to zero if we have:
\begin{equation}
n_r >\frac{\log \frac{\mathrm{Vol}(\mathcal{S}_{A})}{\mathrm{Vol}(\mathcal{S}_u)} }{\log \left( 1 + \frac{\mathrm{Vol}(\mathcal{S}_{A})}{\mathrm{Vol}(\mathcal{S}_{\epsilon}(\mathcal{M}_s))} \right)},
\end{equation}
and subsequently, we have:
\begin{equation}
\mathrm{Vol}(\mathcal{S}_u) > \mathrm{Vol} \left({\mathcal{S}_{A}}\right) 2^{-n_r \log_{2}\left({1+\frac{ \mathrm{Vol}(\mathcal{S}_{A})}{\mathrm{Vol}\left({\mathcal{S}_{\epsilon}(\mathcal{M}_s)}\right)}}\right)}.
\end{equation}
This completes the proof of this theorem.
\color{black}
\ifCLASSOPTIONcaptionsoff
  \newpage
\fi
\bibliographystyle{IEEEtran}
\bibliography{references}
%\IEEEtriggeratref{6}
% that's all folks

\end{document}